\documentclass[%
reprint,
superscriptaddress,
 amsmath,amssymb,
 aps,
]{revtex4-1}

\usepackage{amsmath}
\usepackage{algorithm}
\usepackage[noend]{algpseudocode}

\makeatletter
\def\BState{\State\hskip-\ALG@thistlm}
\makeatother

\usepackage{graphicx}
\usepackage{dcolumn}
\usepackage{bm}
\usepackage[flushleft]{threeparttable} 

\usepackage{dsfont}
\usepackage{color,psfrag}
\usepackage{mathtools,amscd}

\usepackage{soul}
\usepackage{xcolor} 
\definecolor{purple}{rgb}{0.5,0.0,0.5}
\definecolor{mypink1}{RGB}{219, 48, 122}
\definecolor{mypink2}{cmyk}{0, 0.7808, 0.4429, 0.1412}
\definecolor{mygray}{gray}{0.6}

\newcommand\here[1]{\fcolorbox{red}{red}{\rule{0pt}{6pt}\rule{6pt}{0pt}}\quad}

\usepackage{booktabs}
\usepackage{harpoon}

\graphicspath{{figures/}}
\usepackage{pgfplots}
\pgfplotsset{compat=1.12}
\usepackage{xcolor}
\usepackage{tikz}

\definecolor{ForestGreen}{RGB}{34,139,34}

\usetikzlibrary{graphs, calc, automata, graphs.standard, shapes.misc, positioning, fit, shadows, calc, shapes, patterns, arrows.meta, external, pgfplots.fillbetween, intersections}
\usetikzlibrary{decorations.pathmorphing, fadings, shadings}
\pgfplotsset{width=10cm,compat=newest}
\usepgfplotslibrary{units}
\usetikzlibrary{spy,backgrounds}

\usepackage{ifpdf}
    \ifpdf
\usepackage{tikz}
\usetikzlibrary{arrows,chains,matrix,positioning,scopes, fit, calc}
\usetikzlibrary{cd}
\usepackage{chemfig}
\fi
\usepackage{hyperref}

\hypersetup{
	colorlinks,
	linkcolor=blue,
	citecolor=blue, 
	filecolor=black,
	urlcolor=blue}

\begin{document}

\preprint{APS/123-QED}

\title{Super-resolution in Molecular Dynamics Trajectory Reconstruction with Bi-Directional Neural Networks}

\author{Ludwig Winkler}
\affiliation{%
 Machine Learning Group, Technische Universit\"at Berlin, 10587 Berlin, Germany
}%
\author{Klaus-Robert M\"uller}%
 \email{klaus-robert.mueller@tu-berlin.de}
\affiliation{%
 Machine Learning Group, Technische Universit\"at Berlin, 10587 Berlin, Germany
}
\affiliation{%
 BIFOLD --  Berlin Institute for the Foundations of Learning and Data, Berlin, Germany
}
\affiliation{%
Department of Artificial Intelligence, Korea University, Anam-dong, Seongbuk-gu, Seoul 136-713, Korea
}%
\affiliation{%
Max Planck Institute for Informatics, Stuhlsatzenhausweg, 66123 Saarbr\"ucken, Germany}
\affiliation{%
Google Research, Brain team, Berlin, Germany
}
\author{Huziel E. Sauceda}
\email{sauceda@tu-berlin.de}
\affiliation{%
 Machine Learning Group, Technische Universit\"at Berlin, 10587 Berlin, Germany
}%
\affiliation{%
 BASLEARN, BASF-TU joint Lab, Technische Universit\"at Berlin, 10587 Berlin, Germany
}%
\affiliation{Instituto de F\'isica, Universidad Nacional Aut\'onoma de M\'exico,
Apartado Postal 20-346, 01000 M\'exico, D.F., M\'exico}
\date{\today}

\begin{abstract}
Molecular dynamics simulations are a cornerstone in science, allowing to investigate from the system's thermodynamics to analyse intricate molecular interactions.
In general, to create extended molecular trajectories can be a computationally expensive process, for example, when running \textit{ab-initio} simulations.
Hence, repeating such calculations to either obtain more accurate thermodynamics or to get a higher resolution in the dynamics generated by a fine-grained quantum interaction can be time- and computationally-consuming. In this work, we explore different machine learning (ML) methodologies to increase the resolution of molecular dynamics trajectories on-demand within a post-processing step. 
As a proof of concept, we analyse the performance of bi-directional neural networks such as neural ODEs, Hamiltonian networks, recurrent neural networks and LSTMs, as well as the uni-directional variants as a reference, for   molecular dynamics simulations (here: the MD17 dataset).
We have found that Bi-LSTMs are the best performing models; by utilizing the local time-symmetry of thermostated trajectories they can even learn long-range correlations and display high robustness to noisy dynamics across molecular complexity.
Our models can reach accuracies of up to 10$^{-4}$ \AA \ in trajectory interpolation, while faithfully reconstructing several full cycles of unseen intricate high-frequency molecular vibrations, rendering the comparison between the learned and reference trajectories indistinguishable.
The results reported in this work can serve (1) as a baseline for larger systems, as well as (2) for the construction of better MD integrators.
\end{abstract}

\pacs{Valid PACS appear here}

\maketitle

\section{Introduction}

Computational physics and in particular molecular dynamics (MD) simulations are fundamental tools for the understanding of molecular systems.~\cite{tuckerman2010statistical,noe2020machine,Noe2021_JPCLett}
According to the ergodic theorem, the statistical information embedded in the MD generated trajectories is equivalent to performing thermodynamical ensemble averages.~\cite{tuckerman2010statistical}
This is a valid statement for very long molecular  trajectories, meaning that generating predictive thermodynamical observables using \textit{ab initio} methodologies will demand large computational resources.
In order to reduce the computational load, a number of simulation parameters can be tuned to considerably reduce the amount of costly electronic structure calculations.
For example, it is known that performing trustworthy thermodynamical averages from MD results requires the use of non-correlated samples from the trajectory, which in turn allows dismissing intermediate steps in the simulations.
This means that, if the MD integration time step is $dt$, then we can store samples every $n$ step (i.e. $d\tau =n*dt$).
Nevertheless, by doing so, valuable dynamical information is being dismissed, for example subtle dynamical processes and fine correlations in the system.

The parameter $dt$ in MD simulations represent a parameter that has to be chosen in a way to maximize computational efficiency while keeping the precision required to generate physically meaningful results.
On one hand, the integration step $dt$ has to be small enough to generate energy conservative dynamics and to accurately sample high frequency modes.
On the other hand, it also has to be large enough to reduce unnecessary use of computational resources, which then could also lead to non-smooth trajectories and important features could be overlooked. 

Beyond the well known techniques developed within the field of statistical thermodynamics to discretize and integrate Newton's equations of motion~\cite{tuckerman2010statistical}, in recent years, innovative machine learning methods have appeared to accelerate and improve MD simulations ranging from force fields learning~\cite{Behler.ACIE56.2017,SchNetNIPS2017,SchNet2018,Han_DeeP_2018,gdml,sgdml,Shapeev_Moment_MLSciT2021,Bartok_GAP_PRL2010,Zeni_De_Vita2018,Glielmo.PRB97.2018,unke2020machine,von2020exploring,Tkatchenko2020NatCommun,noe2020machine,musil2021physics,keith2021combining,unke2021spookynet,Sauceda_bigdml,li2021graphNNFF,sauceda_NQE_NatComm2021} to avoid expensive \textit{ab initio} evaluations, direct free energy sampling techniques~\cite{tsai2020LMMD,thaler_traj_2021,vlachas2021multiscale,vlachas2021accelerated} as well as new integrator learning~\cite{kadupitiya2020simulating,chen2018neural,greydanus2019hamiltonian,cranmer2020lagrangian}. Additionally, extensive software has been developed coupling modern machine learning models with molecular dynamics techniques~\cite{sGDMLsoftware2019,Doerr_TorchMD_JCTC2021,SchNetPack}.
Each one of those approaches contributes to a different aspect of the task of accelerating molecular simulations, notably with the learned integrator of equations of motion the least studied up to now.
This last area of trajectory and integrators' learning is precisely the topic that we will address in this article.

In this regard, some early attempts to replace conventional dynamic evolution methodologies were developed first by introducing the concept of neural ordinary differential equations (NODEs)~\cite{chen2018neural}, and then by modifying deep neural network (NN) architectures to include physical inductive biases.
For example, Greydanus et al.~\cite{greydanus2019hamiltonian} introduced Hamiltonian NNs (HNN) which compute the velocity and acceleration of a dynamical system by approximating the Hamiltonian in classical mechanics with a NN and subsequently computing the partial derivatives of the approximated Hamiltonian, $\dot{q}=\partial H/\partial p$ and $\dot{p}=-\partial H/\partial q$.
Then following this idea, Cranmer et al.~\cite{cranmer2020lagrangian} expanded on such approach by approximating the Lagrangian of the system introducing Lagrangian NNs (LNN), which relies on computing the partial derivatives corresponding to velocity and acceleration with the Lagrangian formalism.
The architectural configuration of HNN's and LNN's endow the models with the principle of energy conservation.
Even though these methods give good results for ideal systems, their energy conserving bias is then a limitation when describing realistic systems such as thermostated molecular simulations.

Recently, more robust methodologies have been applied to overcome some limitations of analytically biased models, for example, Tsai et al.~\cite{tsai2020LMMD} used modern natural language processing models to predict the probability distributions of dynamical systems allowing the direct calculation of thermodynamical observables.
Another important example was presented by Kadupitiya et al.~\cite{kadupitiya2020simulating} where by means of uni-directional recurrent neural networks (RNN) they manage to forecast the positions and velocities for low-dimensional systems.
The methodologies above-mentioned, even though they have made considerable advances in the field of MD simulations, are still missing a robust analysis of their applicability on realistic scenarios and their interpolation accuracy.

In this article, we introduce a series of approaches to create time-super-resolution of MD trajectories using a wide variety of NN based methods.
In particular, we focus on bi-directional recurrent neural networks (LSTM) to precisely address this issue in the formalism of learnable differential equations.
Our method is used to interpolate missing trajectory snapshots, thereby increasing the resolution of the simulations in a post-processing manner.
Hence, it allows to enhance on-demand the resolution of dynamical properties such as vibrational spectra, better free energy surfaces, as well as more detailed animations.
Additionally, from applying our methodology to realistic molecular systems, we have found an insightful trend (which may not come as a complete surprise): the higher the temperature of the system, the easier to learn its dynamics is. 
This result supports the hypothesis that the Helmholtz free energy is less complex than the underlying potential energy surface. 
In other words, as the temperature increases, the dynamics of the system becomes smoother.
Which means that by combining MD simulations with our post-processing super-resolution method, the computational cost as well as simulations times could be considerably reduced.

\begin{figure}[t]
\centering
\includegraphics[width=0.9\columnwidth]{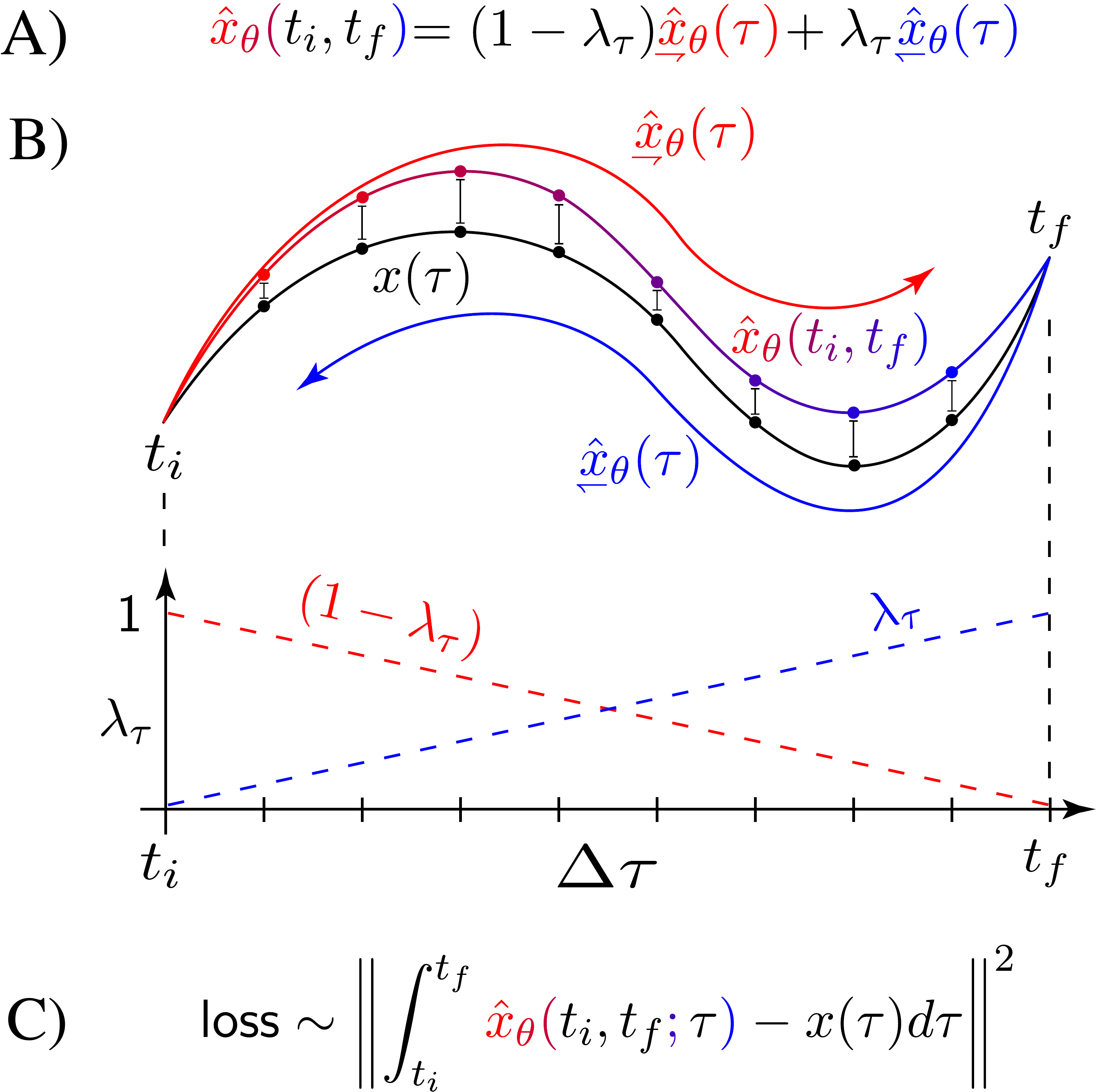}
\caption{Interpolation of the trajectory using a bi-directional NNs. 
A) Linear combination of the forward $\overrightharp{x}_\tau$ (red) and backward $\overleftharp{x}_\tau$ (blue) solutions allows for better control of the error by using the information of both the initial and the final condition, $x(t_i)$ and $x(t_f)$, respectively.
$\lambda_\tau$ is the mixing parameter defined by eq.~\ref{eq:lambda}. 
B) Graphical description of the interpolation process by combining the forward and backward predictors.
C) Loss function.}
\label{fig:bi_interpol}
\end{figure}



\section{Methodology}

\subsection{Molecular dynamics}
Performing MD simulations of a molecular system in practice requires discretizing in time Newtons' equations of motion $\ddot{\mathbf{a}}=m^{-1}{\mathbf{F}}$ using, for example, the Velocity-Verlet algorithm given a potential energy surface (PES) $U=U(\mathbf{r})$ which defines the force field $\mathbf{F}=-\nabla_{\mathbf{r}} U$ acting on each atom. 
Hence, as a result, we obtain a trajectory with the molecular time evolution or in other words, a time series of the atomic coordinates \textbf{r} and momentum \textbf{p}: $\mathcal{S}=\{ \mathbf{x}_t=(\mathbf{r}_t,\mathbf{p}_t); t=i\Delta t, i=0,\ldots,N_T \}$. Here $\mathbf{x}$ is known as a point in the phase space of the system.
The discretization parameter $dt$ has to be selected according to the system and the simulations conditions, being for example a key aspect the fact that has to be much smaller than the fastest oscillation period in the system.
Which, due to the nature of chemical bonds, renders $dt$ in the order of 0.1 to 1 fs for organic systems.
Now, another fundamental aspect to generate meaningful simulations is the length of the time series or the total simulation time $T=N_T*dt$, which contrasting the selection of $dt$, it has to be much larger than the slowest oscillation in the system.
According to the ergodic theorem, in order to fully recover the thermodynamical properties of the system $N_T\to\infty$, nevertheless in practice a careful selection of a finite value of these parameters gives accurate results.
Thereby, the selection of $N_T$ is a more abstract task because it is tightly correlated to the type of system and the physical phenomena to be studied.
For example, studying light diatomic molecules could require simulation times on the order of tens of fs ($N_T\approx10^3$), while more interesting molecules containing a couple of dozens of atoms and a fluxional molecular structure would require simulations on the order of ns ($N_T\approx10^7$)~\cite{sGDML_Appl_jcp,Sauceda_bigdml}.
Keeping in mind that every integration step during the MD simulation requires the explicit calculation of the potential energy and forces, which in the case of electronic structure calculations such as Density Functional Theory (DFT) can take on the order of seconds, we can see the benefit on creating a methodology that helps to reduce the value of $N_T$ without losing accuracy.
In a straightforward manner, this implies that we take larger integration steps in the simulation.

\begin{figure*}[hbtp!]
\includegraphics[width=0.92\textwidth]{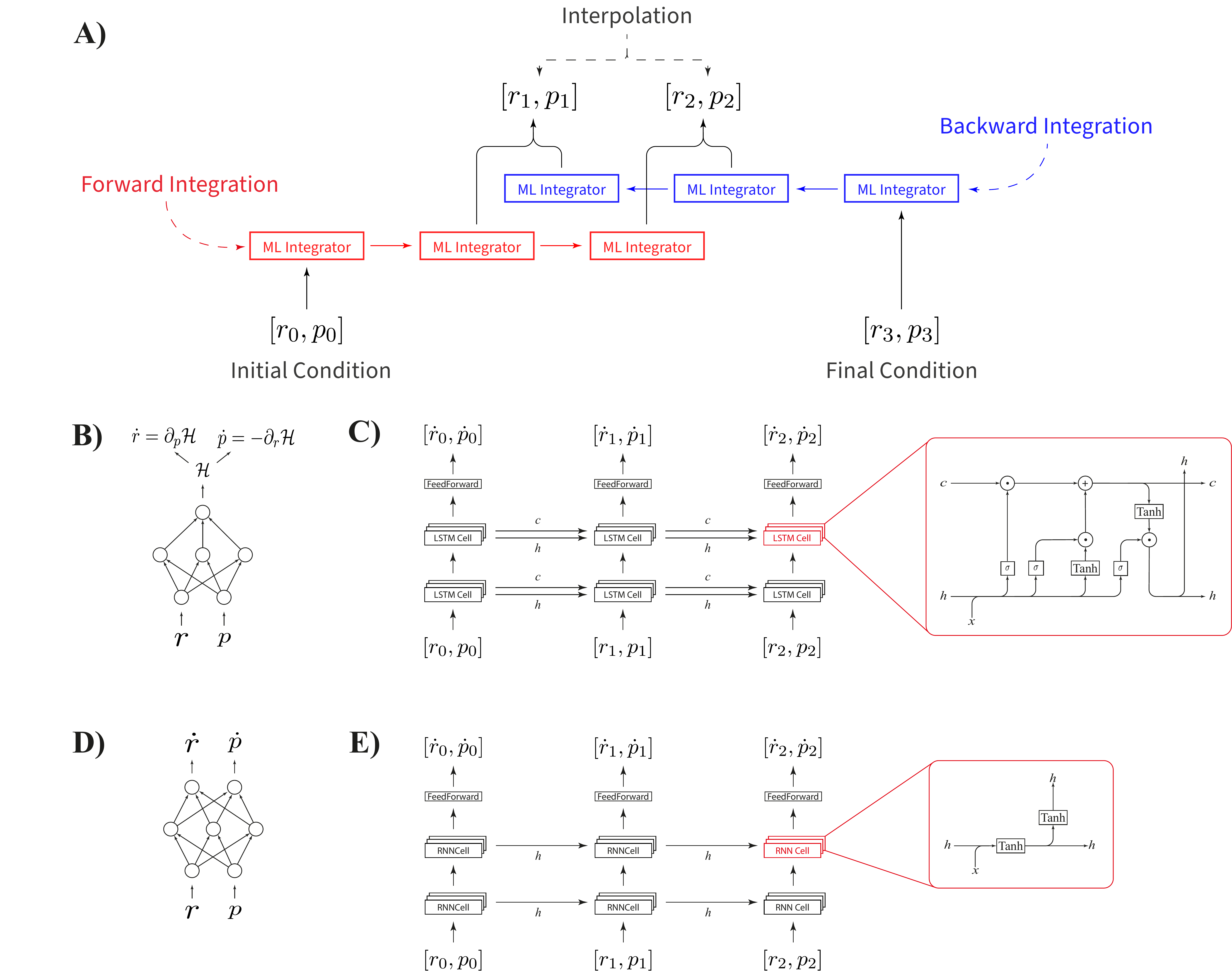}
\caption{A) The information flow in a Bi-directional integrator, which is trained to interpolate the differential equation that governs the molecular dynamics. The same neural network differential equation is utilized to predict the solution from the initial condition $[r_0,p_0]$ at $t=0$ forward in time and from the final condition $[r_3,p_3]$ at $t=3$ backward in time. Subsequently, the forward and backward solution is interpolated with $\lambda_\tau$ to obtain the interpolated values. B) The Hamiltonian Neural Network (HNN) architecture which uses partial derivatives of the network output in the forward pass. C) The information propagation of LSTM neural networks and the inner dynamics of each LSTM cell consists of aggregating the input information $x$ with the memory cell $c$ and computing the output $h$. D) Neural ordinary differential equations use neural networks in black box adaptive ODE solvers. E) RNN model architecture showcasing how the neural network integrators propagate and aggregate information through time in their hidden state $h$.}
\label{fig:LSTM}
\end{figure*}

\subsection{MD Trajectory Interpolation with Neural Networks}

In this article, we propose to employ machine learning algorithms to integrate Newton's laws of motion directly from the phase space vector representation of molecules.
To that end, we train a machine learning integrator to interpolate the phase space trajectory of the molecules over a finite time horizon.
In general, $\mathbf{r}$, $\mathbf{p}\in\mathbb{R}^{3N}$ where $N$ is the number of atoms in the molecular system, but for the sake of simplicity here we will analyse the one dimensional case, $r$, $p \in \mathbb{R}$ and later generalize to $3N$ dimensions.

A time-dependent variable can be described by the differential equation $\dot{x}_t = f(x_t, t)$ for a general time index $t$, which is derived from the phase space vector, which then has the discrete solution,
\begin{align}
    x_{t+\Delta t} = x_t + \int_{t}^{t+\Delta t} f(x_{t'},t') dt'.
\label{eq:ODEsol}
\end{align}

Now, instead of performing a "big" jump by $\Delta t$, we would like to interpolate in between the two subsequent phase space vectors $x_{t+\Delta t}$ and $x_t$ to achieve a higher resolution.
Hence, we choose to integrate a parameterized dynamics described by, $\dot{x}_\tau = f_\theta(x_\tau, \tau)$ which has a higher temporal resolution $\tau$ between any two subsequent states in the simulation $x_t$ (i.e. $\tau\in [t,t+\Delta t]$).
Thereby, two subsequent phase space vectors in the coarse simulation separated by a time $\Delta t$ takes the form,
\begin{align}
    x_{t+\Delta t} = x_t + \int_{t}^{t+\Delta t} f_\theta(x_\tau, \tau) d\tau
\label{eq:ODEsol2}
\end{align}

\noindent where the second term is the neural network integrator that propagates the dynamics.
To infer the correct set of parameters $\theta$ of the integrator network,
we have to optimize the scalar loss function, 
\begin{align}
\label{loss}
    &\ \quad \mathcal{L}(f_\theta(x_\tau, \tau), f(x_\tau, \tau), t, t+\Delta t)= \\
    & \frac{1}{\Delta t} \left\lVert \int_{t}^{t+\Delta t} \bigg[ f_\theta(x_\tau, \tau) - f(x_\tau, \tau) \bigg] d\tau \right\rVert^2.
\end{align}

The loss function is evaluated on randomly sampled sections of the trajectories in mini batches.

\subsubsection{Bi-Directional Interpolation}

The coarse simulation $x_t$ provides a trajectory at discrete time steps $\Delta t$, from which the learned 
intermediate dynamics $f_\theta(x_\tau, \tau)$ reconstructs the missing trajectory segment at a higher resolution (See Fig.~\ref{fig:bi_interpol}-B). 
This means obtaining $x_\tau$, where $\tau \in \{ t, t+\Delta\tau,\dots, t+(n-1)\Delta\tau,t+\Delta t \}$ and $\Delta\tau/\Delta t$ gives the enhanced resolution.
The coarse trajectory at time steps $x_t$ and $x_{t+\Delta t}$ thus provides the initial and the final condition for the dynamics $f_\theta(x_\tau, \tau)$.

Now, for better error control we can make use of the fundamental property of time reversibility of Newtons' equations, meaning that we obtain the same trajectory if we start from the initial conditions $x_0$ and get to the final state $x_T$ and if we start from $x_T$ and propagate the system backwards in time to $x_0$. 
With the provision of the initial condition $x_t$ and the final condition $x_{t+\Delta t}$, we can compute the forward $\overrightharp{x}_{\tau}$ and backward $\overleftharp{x}_\tau$ trajectories for $\tau \in \{ t, t+ \Delta t \}$ (see Figs.~\ref{fig:bi_interpol}),

\begin{align}
    \overrightharp{x}_{\tau} &= x_t + \int_{t}^{\tau} f_\theta (x_s, s) ds \\ 
    \overleftharp{x}_{\tau} &= x_{t+\Delta t} - \int_{t+\Delta t}^{\tau} f_\theta (x_s, s) ds
\end{align}

The interpolation of the two trajectories is achieved via the interpolation parameter $\lambda_\tau \in [0,1], \tau \in [0, \Delta t]$,

\begin{align}\label{eq:lambda}
    \lambda_\tau = \frac{\int_{s=0}^{\tau} s \ ds}{\Delta t}
\end{align}

\noindent which is a monotonically increasing for the duration $\tau \in [0, \Delta t]$ and is designed to shift the weight from the forward trajectory to the backward trajectory over the course of the interpolation.
The interpolated trajectory can then be recovered via,

\begin{align}
    x_\tau = (1 - \lambda_\tau) \overrightharp{x}_\tau + \lambda_\tau \overleftharp{x}_\tau.
\end{align}

The interpolation is visualized in Fig. \ref{fig:bi_interpol}, which highlights the shifting interpolation parameter $\lambda_\tau$.

\subsubsection{Neural Network Integrators}

In order to assess which NN architecture is the most suitable to interpolate MD trajectories, we considered four NN integrators: neural ODEs, Hamiltonian neural networks, recurrent neural networks and LSTM. A pictorial description of the considered architectures is presented in Fig.~\ref{fig:LSTM}.

\noindent\textbf{Differential architectures}: 
The most accessible approach to modeling the approximate dynamics $f_\theta$ is by directly computing the differentials with a fully-connected neural network.
The Euler discretization is often sufficient for convergence but more sophisticated, adaptive solvers are applicable such as Runge-Kutta and Dormand-Prince solvers for differential equations \cite{dormand1980family}.
In order to be applicable to deep neural networks trained with the backpropagation algorithms, these solvers require the use of the adjoint sensitivity method, which backpropagates an adjoint quantity as a surrogate gradient through time \cite{chen2018neural}.
The adjoint sensitivity method backpropagates the error through adaptive solvers with a constant memory cost which is highly suitable for adaptive solvers which potentially large number of evaluations.
Once the adjoint is backpropagated to all evaluations, the gradients of the parameters $\theta$ can be obtained and gradient descent training is eligible.\\

Newton's equation for dynamical systems can be generalized into the Hamiltonian mechanics framework.
The canonical coordinate positions $r_t$ and momentum $p_t$ can be obtained through the partial derivatives of  the Hamiltonian $\mathcal{H}(r_t, p_t)$.
Hamiltonian neural networks~\cite{greydanus2019hamiltonian} predict an approximate Hamiltonian $\mathcal{H}_\theta \approx \mathcal{H}$ parameterized as a deep neural network and compute the time derivatives $\dot{r}_t$ and $\dot{p}_t$ during the forward pass.
In terms of functional analysis, one has to be careful in using an architecture which is differentiable at least twice, since the network is trained through backpropagation which requires a second differentiation of the model.
In practice, this amounts to using continuously differentiable activation functions such as tangent hyperbolic or sigmoid and refrain from using piece-wise linear activation functions such as rectified linear units.
By virtue of their construction, Hamiltonian networks exhibit energy conserving properties, such that the total energy of a dynamical system remains constant.
This is of interest for energy-based systems such as molecular dynamics, in which energy is shifted between potential and kinetic energy but never lost.\\

\noindent\textbf{Recurrent architectures}: 
RNNs extend feed-forward neural networks through recurrent connections through time.
They offer the ability to explicitly model time dependent relationships by incorporating the neuron activations of the previous time step.
Long Short-Term Memory (LSTM) networks are recurrent architectures that resolve some important issues on general RNNs by using memory cells that can selectively read and write to them.
LSTM's are widely used in modelling time series and provide a remedy for the vanishing and exploding gradient problem of classical RNN's due to excessive or miniscule eigenvalues in the recurrent weight matrices \cite{pascanu2013difficulty}.

Whereas Ordinary Differential Networks and Hamiltonian Networks are considered Markovian in the sense that they only use the current state $x_t$ of the dynamical system to predict the time derivative $\dot{x}_t$, RNNs and LSTMs are capable of modelling long-distance dependencies through their hidden states in memory cells.
In the next section, we analyse in detail the performance of these two different approaches, as well as their bi-directional variants on the task of trajectory reconstruction.

\begin{figure*}[ht!]
\includegraphics[width=1.0\textwidth]{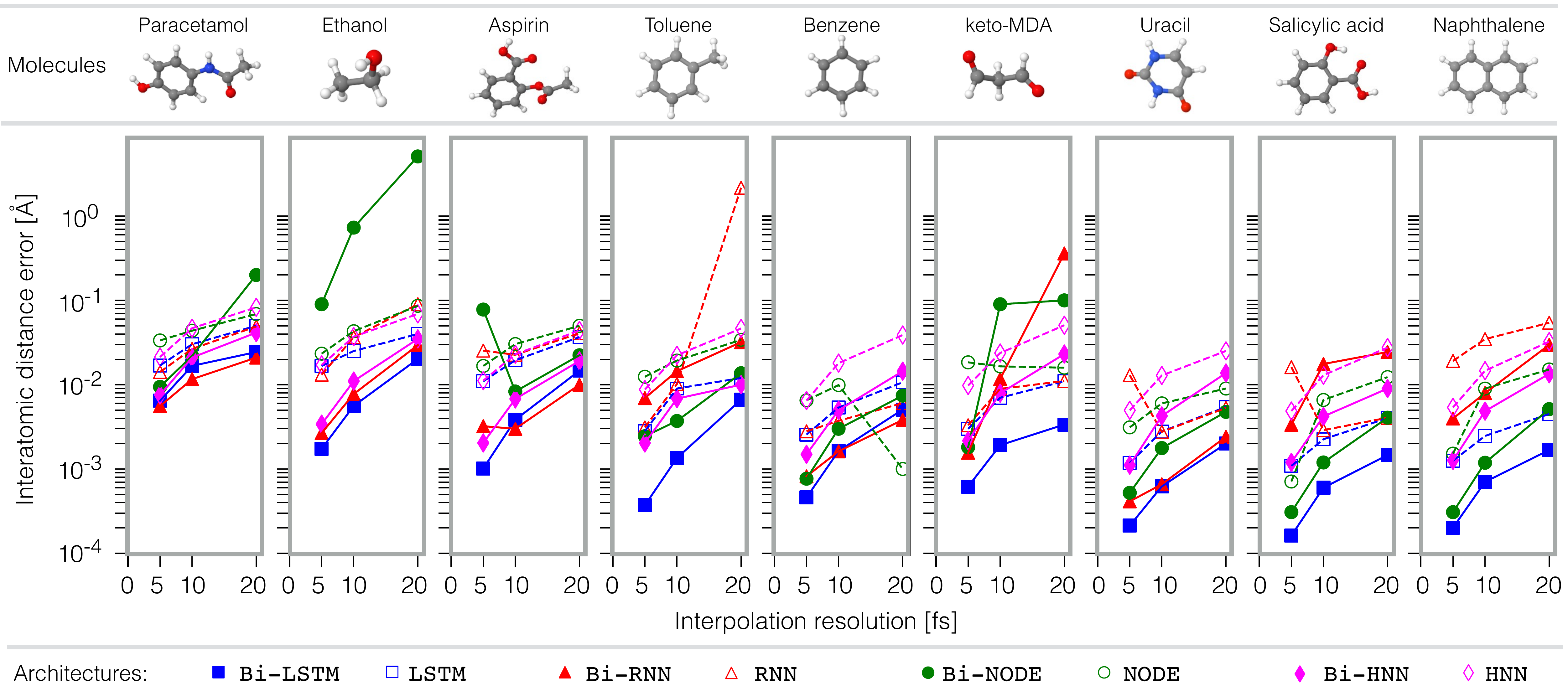}
\caption{Performance comparison of uni- and bi-directional models over multiple integration lengths $\Delta t$ on the extended-MD17 dataset~\cite{gdml,sGDML_Appl_jcp}. Uni-directional and bi-directional methods are represented by empty and full symbols, respectively. The considered interpolation resolutions are 5, 10, and 20 in femtoseconds.}
\label{fig:performance}
\end{figure*}

\begin{figure*}[hbtp!]
\includegraphics[width=\textwidth]{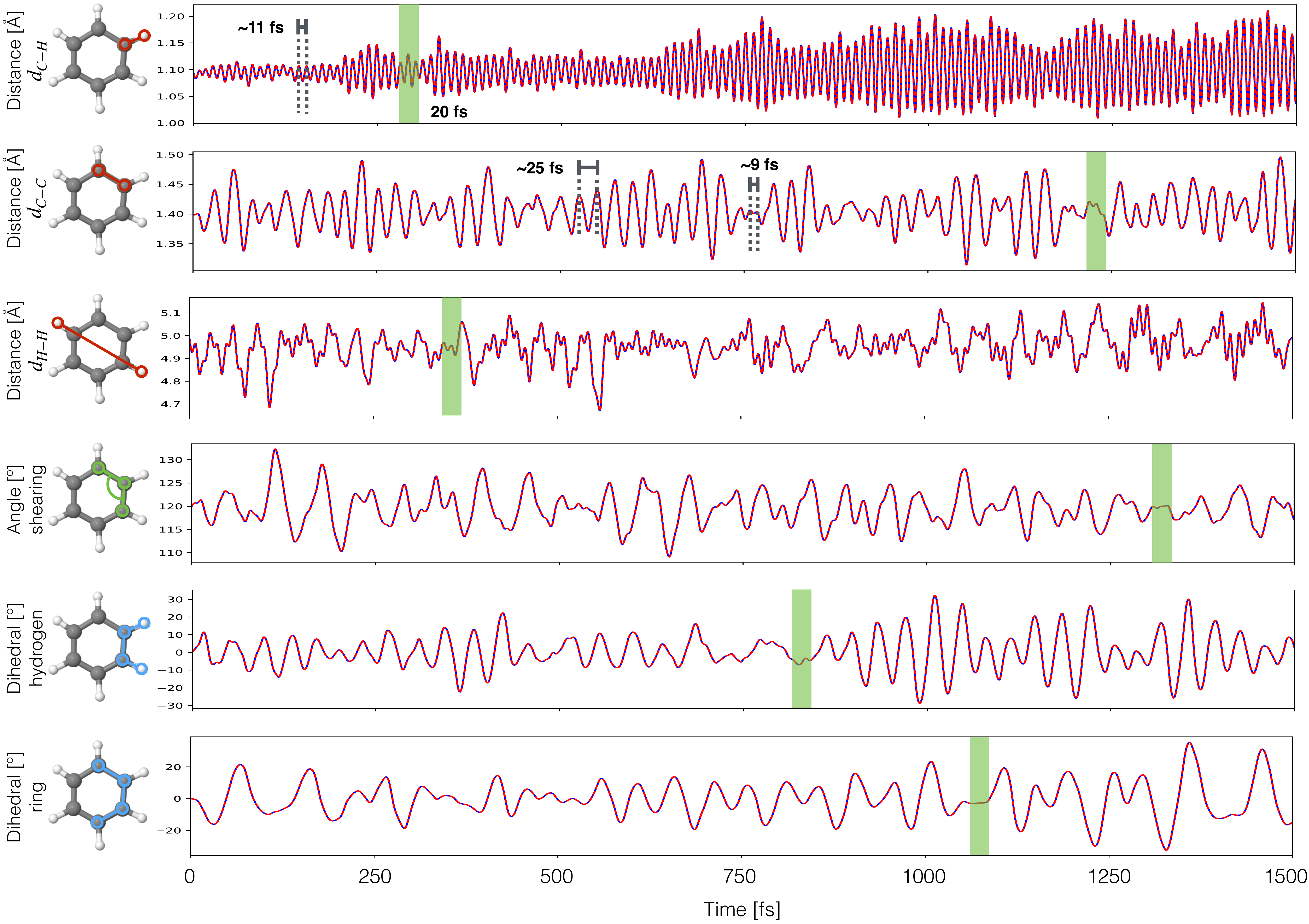}
\caption{Analysis of the benzene dynamics and its interpolation for the 20 fs interpolation case. The six internal variables displayed are representative feature of the molecule, including interatomic distances (in \AA), internal angles (in degrees), and dihedral angles (in degrees) (represented in red, green and blue in the left-most molecular structure, respectively). The true dynamics is displayed by the blue curve and the predicted results are given by the dashed red lines. In grey dotted lines we show some relevant periods for some variables, and in a green window we represent the 20 fs interpolation interval.}
\label{fig:trajBenzene}
\end{figure*}

\section{Results and Discussion}

\subsection{Training and validation of the model}
The general goal of this study is to stablish the applicability of the presented methodologies on realistic dynamics from physico-chemical simulations.
To this end, we have selected the well established extended-MD17 dataset~\cite{gdml,sGDML_Appl_jcp} which contains middle-sized molecules with various dynamical complexities.
In this section, we analyse the performance of the bi-directional approaches (i.e. LSTM, RNN, HNN, and NODE) compared to their uni-directional counterparts while considering different resolution for the interpolation tasks. 
These results are shown in Fig.~\ref{fig:performance}.

\subsubsection{Uni- vs Bi-directional NNs}
Common methods for sequence and trajectory learning are usually deployed using the natural arrow of time, nevertheless it has been demonstrated in other areas of machine learning, such as language modelling~\cite{vaswani2017attention, bahdanau2014neural, GoogleNMT}, that bi-directional learning substantially improve the results.
Here, we exploit the regularity of the physical trajectories and local time reversal symmetry by using bi-directional versions of the architectures shown in Fig.~\ref{fig:LSTM}.
In order to directly show the benefit of this approach, in Fig.~\ref{fig:performance} we compare the performance of the two approaches for all the molecules from the extended-MD17 dataset~\cite{gdml}.
From this figure, we can quickly see that using bi-directional NNs can boost the accuracy of the model by up to two orders of magnitude.
This reveals that including both directions of the arrow of time goes beyond simply duplicating the amount of data, instead this strongly encodes the trajectory's regularity in the model which results in a more accurate description without increasing the number of parameters in the architecture.
Additionally, we can see that this phenomenon is not architecture specific, since all the tested NNs obtained a similar increase in accuracy and similar learning curves when considering bi-directionality.
Additionally, such behaviour is also independent of the interpolation resolution.

\subsubsection{Bi-LSTM performance}
Now, comparing the different bi-directional methods, we can see that recurrent neural networks (i.e. RNN and LSTM) always perform better than neural ODE based approaches.
This is due to the fact that recurrent based NNs are more robust against noise, as will be the case when dealing with thermostated molecular dynamics simulations.
Furthermore, Fig.~\ref{fig:performance} shows that the Bi-LSTM architecture (see Fig.~\ref{fig:LSTM}) gives more consistent results across molecular structures with different complexities compared to the rest of the architectures, reaching interpolation errors of up to one order of magnitude lower. 
This advantage of Bi-LSTMs over other architectures could be related to the fact that LSTM cells propagate information more efficiently and keep track of the many subtleties in the molecular trajectories for longer periods of time.
It is worth to remark that the accuracies reached by Bi-LSTMs in real space (i.e. coordinates) range on the order of 10$^{-3}$ \AA, which in practical terms is indistinguishable from the reference trajectory.
Given these results, from now on, we will focus our analysis only on Bi-LSTM networks.

\subsubsection{Molecular complexity and degrees of freedom}
As an additional observation, we have noticed that molecules with higher complexity (e.g. larger number of rotors and/or higher fluxionality) generate more intricate trajectories that render the reconstruction process more challenging.
This is only evident for the cases of paracetamol, ethanol and aspirin, nevertheless, there are cases where the correlation between interpolation accuracy and molecular morphology is non-intuitive. 
For example, even though the keto-MDA molecule (see Fig.~\ref{fig:performance}) presents a complex PES with two rotors as main degrees of freedom~\cite{sgdml_bookAppl}, its trajectory's reconstruction precision is similar to benzene and uracil, molecules that have no rotors and are formed only by an aromatic ring.
Furthermore, the best accuracy reached by the Bi-LSTM model was for salicylic acid, a molecule that has two rotors coupled by a complex hydrogen bond where the proton is dynamically being transferred between the two functional groups.

The physical interpretation of these results is that the reconstruction process occurs on the free energy surface (FES) of the molecular system, which means that the molecule is moving on its FES instead of the PES.
Hence, moving on the molecular FES at a given temperature can result on non-trivial behaviour originated from the entropic contributions. 
Thereby, an apparently intricate molecular system such as salicylic acid can generate a smother FES that uracil or naphthalene (see Fig.~\ref{fig:performance} for molecular structure reference).
The insight behind these results is that, for the considered molecules,  thermal fluctuations considerably reduce the dynamical complexity of molecular systems with high fluxionality. Thereby, simplifying the learning process.

\subsection{Achieving super-resolution in MD trajectories}

In the previous section, we have demonstrated that Bi-LSTMs are a suitable architecture for trajectory learning of realistic molecular simulations.
In this section, we continue analysing what it implies to reconstruct molecular trajectories, but now from the vibrational point of view, first from the normal frequencies framework and then from the real dynamics perspective.

One of the keys aspects in trajectory interpolation is the time resolution that can be achieved by the model.
Given that the MD17 dataset has integration steps of 1 fs, here the task was to skip a number of frames $n$ for each molecule and then reconstruct the missing fragments on the trajectory.
Hence, for this dataset, $\Delta \tau=1$ fs and $\Delta t=n$ fs.
In this regard, Fig.~\ref{fig:performance} shows the performance of all the models for $\Delta t=$ 5, 10 and 20 fs.
Let's analyse the results for each value of $\Delta t$ in the context of its physical implications for the vibrational normal modes (i.e. harmonic analysis).
As a reference, the fastest atomic oscillation periods in paracetamol and in benzene are 
$\approx$9.3 fs and $\approx$11 fs, respectively~\cite{sGDMLsoftware2019}.
Such vibrations are mainly due to fast oscillations of the hydrogen atoms in the molecule.
Hence, sampling a trajectory every 5 fs in the context of molecular vibrations means that we are skipping half the period of the fastest oscillation in the molecule, which implies that, at most, the model has to interpolate half of the oscillation cycle.
This fact is reflected on the validations errors shown in Fig.~\ref{fig:performance}.
More challenging cases are sampling molecular trajectories every 10 and 20 fs, given that in these two cases the model has to reconstruct at least one full period (in the case of $\Delta t=$10 fs).
Furthermore, the most interesting resolution to analyse in more detail is $\Delta t=$20 fs, given that in such case the model has to reconstruct at least one full cycle of all the molecular oscillations with frequencies larger than $\approx$1600 cm$^{-1}$.
Which in the case of the benzene molecule, there are eight out of 30 normal modes with frequency values larger than such value (i.e. $\sim$27\% of the normal modes).
More interestingly, six of them have oscillation periods of $\approx$10 fs, meaning that the $\Delta t=$20 fs model has to reconstruct two full oscillations periods for the six vibrational modes using as inputs only the initial and final states (i.e. ($\mathbf{r}_t,\mathbf{p}_t$) and ($\mathbf{r}_{t+21},\mathbf{p}_{t+21}$)).
In the Supporting Information, we have animated such example for the toluene molecule case, where the true dynamics is represented by blue atoms and the interpolated dynamics appears in red.
As expected from Fig.~\ref{fig:performance}, the dynamics is practically indistinguishable.

In order to visualize this analysis, in Fig.~\ref{fig:trajBenzene} we present the reconstructed trajectory for the benzene molecule using the Bi-LSTM architecture for $\Delta t=$20 fs, and we show its dynamics in terms of its main internal degrees of freedom: interatomic distances, angles and dihedral angles.
In this figure, the blue curve is the ground truth and the red dashed line represents the predicted dynamics.
A quick glance over the plot, shows that, even though the $\Delta t=$20 fs case in Fig.~\ref{fig:performance} is the one that gives the largest error, \textit{such accuracy still corresponds to an excellent agreement between the reference trajectory and the ML prediction.}

As mentioned before, the fastest oscillation in a molecule are due to hydrogen atoms oscillations, which in the benzene case it can be tracked by plotting the interatomic distance $d_{\text{C-H}}$ (Fig.~\ref{fig:trajBenzene} top panel).
Here, the measured oscillation period from the signal is $\approx$11 fs, slightly larger than the normal mode value.
The origin of this red shift is the fact that at finite temperatures the system visits the anharmonic region of the PES, which then generate slower oscillations.
As a reference, a green rectangular window is used in Fig.~\ref{fig:trajBenzene} to show examples of interpolated intervals of trajectories.
Hence, in the case of $d_{\text{C-H}}$ we can clearly see that the method is successfully reconstructing two full cycles of the variable, rendering indistinguishable the comparison to the reference trajectory. 
Another important internal variable is the first neighbour carbon-carbon distance $d_{\text{C-C}}$ (second panel in Fig.~\ref{fig:trajBenzene}).
$d_{\text{C-C}}$ oscillation period is in general considerably slower, but because such interatomic distance is part of many anharmonically-coupled vibrational modes,
it can create apparent high-frequency diatomic oscillations (as short as $\approx$9 fs).
Here, despite the complex dynamics of this internal variable, the model manages to faithfully recover its behaviour.
Now, in order to incorporate a highly non-linear and weakly correlated interatomic distance, we considered two opposite hydrogen atoms, $d_{\text{H-H}}$ (third panel in Fig.~\ref{fig:trajBenzene}).
In principle, this variable should amplify small errors in the reconstructed trajectory, nevertheless the results are still in excellent agreement.

Another important aspect in molecular fluctuations is the analysis of internal shearing deformations as well as out of plane deformations, which represent a global mechanical property of the system.
These mechanical deformations can be analysed by measuring shearing angles and dihedral angles as shown in the lower half of Fig.~\ref{fig:trajBenzene}.
The oscillations in these variables are considerably slower compared to interatomic distance fluctuations, but they can contain highly anharmonic contributions, making them a good measure for reconstruction accuracy.
Again, here the interpolation accuracy of the model very high, which makes the predicted trajectories indistinguishable from the reference curve.
In fact, the MAE reconstruction accuracy of the Bi-LSTM model for the benzene molecule's trajectory is $\sim$10$^{-4}$~\AA, value that is close to the accuracy of the actual numerical integrators such as the Verlet algorithm.

The results obtained in this section show that the methods presented here, and in particular Bi-LSTMs, give excellent trajectories' reconstruction accuracies, rendering their predictions indistinguishable from full \textit{ab-initio} MD results. 
Hence, this allows to confidently use these techniques in a diverse range of applications such as data augmentation, super-resolution generation, or even storage capacity reduction.
In the next section, we use Bi-LSTMs to extract some physical insights from learning trajectories at different temperatures, where we demonstrate that we can faithfully recover its molecular free energy surface.

\begin{figure}[t]
\centering
\includegraphics[width=\columnwidth]{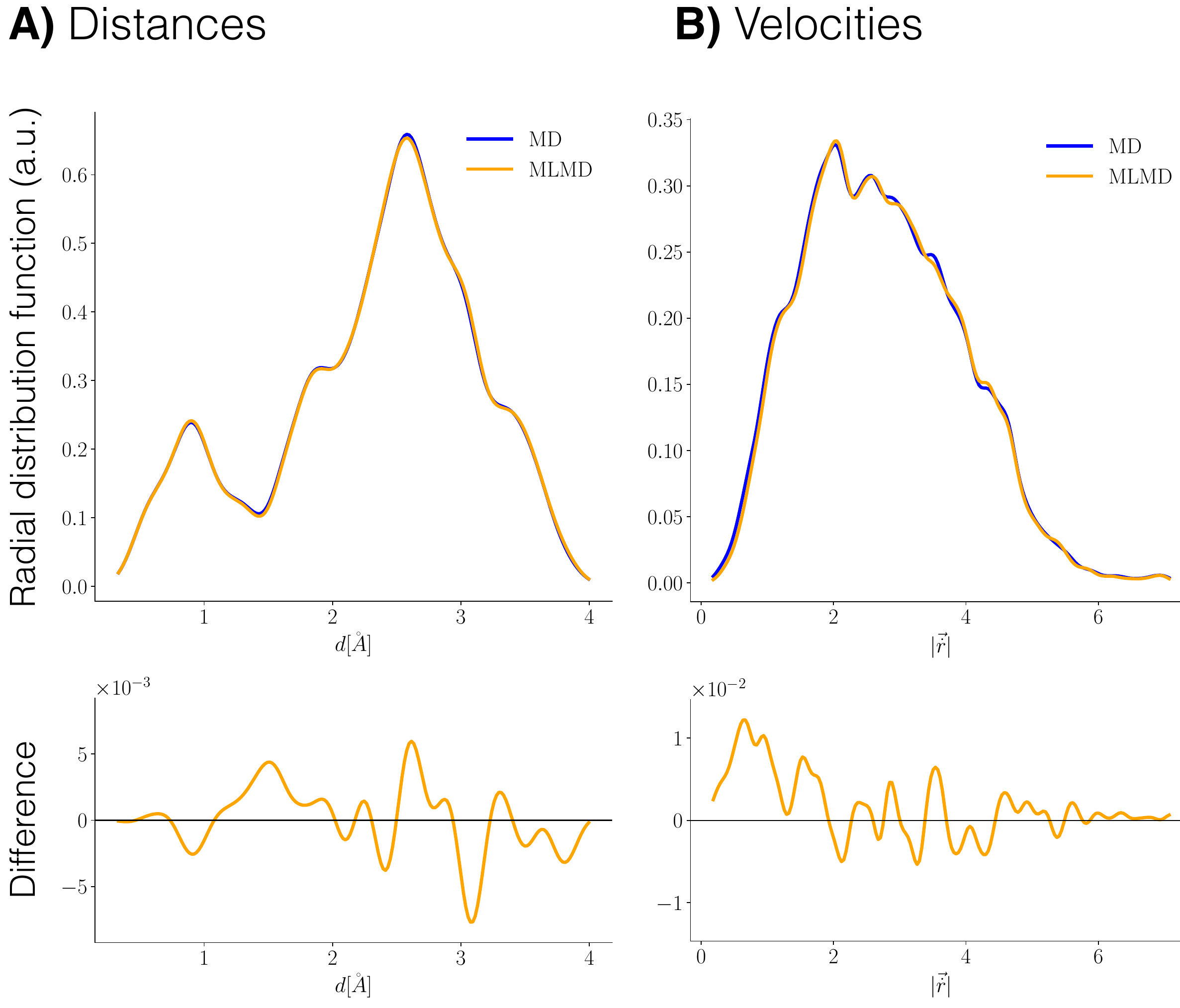}
\caption{Comparison of between the interpolated (yellow) and ground truth (blue) trajectories for keto-MDA molecule. A) Interatomic distance histogram and B) speed histogram. The interpolation was done with a Bi-LSTM integrator with $\Delta t = 10$. The bottom panels provide the difference between the two curves.}
\label{fig:HistMDADistVel}
\end{figure}


\subsection{Physical insights from the reconstruction process: Application to keto-MDA as case of study}

The keto-MDA molecule is a challenging molecule given that it presets a wide variety of interactions that generate a PES with strong electrostatic interactions but also regions with particle-in-a-box behaviour~\cite{sauceda2020JCP,sgdml_bookAppl}.
Hence, such intricate energy landscape generates interesting dynamics from which, for example, force field reconstruction is not a straightforward task and requires the use of sophisticated ML methodologies~\cite{sgdml,SchNet2018,painn,unke2021spookynet}.
Interestingly, there is evidence that the force field learning process gets easier when the temperature of the generated training data gets lower~\cite{sgdml_bookAppl}.

From the statistical point of view, the MD trajectories at a given temperature sample the configuration space in such a way that the molecular free energy surface can be estimated by $\sim \text{ln}~P(\theta_1,\theta_2)$ where $P(\theta_1,\theta_2)$ is the integrated configurational probability density, and $\theta_1$ and $\theta_2$ are the main degrees of freedom of the system (see Fig.~\ref{fig:HistkMDA_distTemps}).
%
In general, the FES is known to get smoother as we increase the temperature due to entropic effects.
This also means that the molecular trajectory in phase space becomes more stable, and hence it should be easier to reconstruct.
In order to get some insights about this effect as well as to further validate the performance of our models, we have run three simulations of the keto-MDA molecule at different temperatures (100K, 300K and 500K) using a pretrained sGDML model~\cite{sGDMLsoftware2019} and analysed the generated trajectories.
In Table~\ref{table:TempTrajLearn_MAEdist_normalized}, we summarize the results for the learning procedure for the three temperatures. 

\begin{table}[!ht]
    \centering
\begin{tabular}{c | *3c}
    \toprule
    \multicolumn{4}{c}{\textbf{A)} Interatomic distances [$10^{-3}$ \AA]}\\
    \cline{1-4}
    Temperature [K] & $T_{\text{tr}}=$5 & $T_{\text{tr}}=$10 & $T_{\text{tr}}=$20        \\
    \cline{1-4}
    100         & 0.259     & 	0.930  &    3.287  \\
    \cline{1-4}
    300         & 0.417    & 	1.500   &   5.321 \\
    \cline{1-4}
    500         & 0.501    & 	1.801   & 	6.399 \\
    \midrule
    \multicolumn{4}{c}{\vspace{0.01cm}} \\
    \toprule
    \multicolumn{4}{c}{\textbf{B)} Velocity field [$10^{-3}$ \AA/fs]}\\
    \cline{1-4}
     Temperature [K] & $T_{\text{tr}}=$5 & $T_{\text{tr}}=$10& $T_{\text{tr}}=$20        \\
    \cline{1-4}
    100         & 11.7100 	 & 38.390 	 & 132.700  	 \\
    \cline{1-4}
    300         & 11.320 	 & 36.820  	 & 127.210 	 \\
    \cline{1-4}
    500        & 10.380 	 & 33.070 	 & 113.60 	 \\
    \midrule
\end{tabular}
	\caption{Experimental results for evaluating the influence of varying training integration time $T_{tr}$ of the Bi-LSTM architecture on A) inter-atomic distances and B) velocity field. This was done for the keto-MDA molecule as a case of study for simulations run at three different temperatures. The interatomic distances and velocity field were normalized to follow the first two moments of a standard Normal distribution.}
	\label{table:TempTrajLearn_MAEdist_normalized}
\end{table}

\subsubsection{Phase-space histograms}
\begin{figure}[t]
\includegraphics[width=\columnwidth]{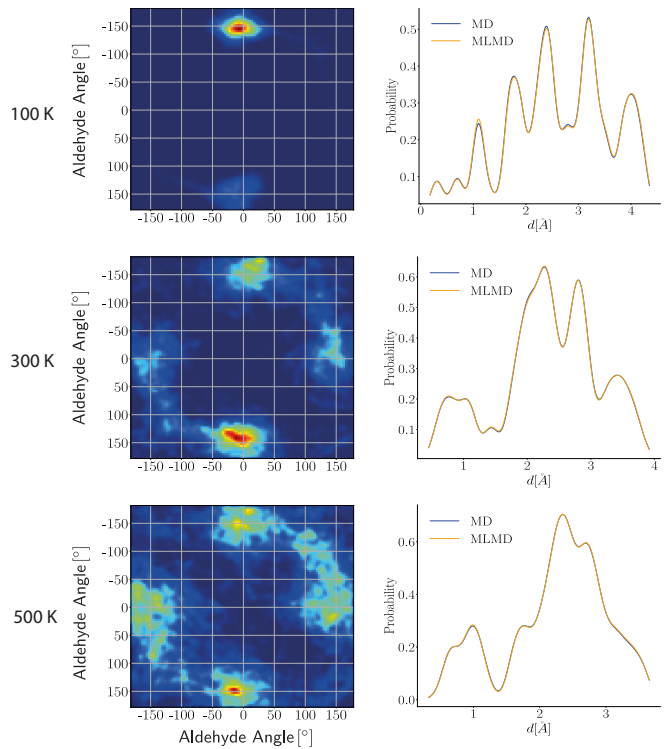}
\caption{The distribution of interatomic distances $d$[\AA] of keto-MDA at 100 K, 300 K and 500 K. The predicted distribution of interatomic distances is shown in red and the target distribution is shown in blue. The distributions become less multi-modal as the temperature increases.
}
\label{fig:HistkMDA_distTemps}
\end{figure}

The frameworks presented in this study work on phase space, $(r,p)$, meaning that during the reconstruction task the trajectory and the velocity field have to be recovered.
Hence, here, we analyse Bi-LSTM's predictive power in terms of physically meaningful distributions from molecular dynamics simulations by examining the speed and interatomic distances distributions.
In previous sections, we have focused on spacial accuracy for trajectory reconstruction (up to $\sim10^{-4}$ \AA), but another important aspect is to recover an accurate velocity field.
In Table~\ref{table:TempTrajLearn_MAEdist_normalized}, we report the interpolation accuracy for both of these variables, displaying an excellent agreement with the reference data.
In particular, in Fig.~\ref{fig:HistMDADistVel} we show the explicit comparison of the reconstructed probability distributions for the interatomic pair distance distribution function, $h(r)$, and the speed distribution function
for the case of keto-MDA's trajectory at 300K.
From figures~\ref{fig:HistMDADistVel} and~\ref{fig:HistkMDA_distTemps} and from Table~\ref{table:TempTrajLearn_MAEdist_normalized}, we can see that the acquired velocity field precision is maintained for the three different temperatures.

\subsubsection{Temperature dependent learning}
It is worth noticing that the temperature has indeed an effect during the learning process. 
From the dynamical point of view, we know that the free energy surface depends on the temperature, meaning that the trajectory generated will follow different patterns and statistically sample differently the phase-space.
This can be seen in Fig.~\ref{fig:HistkMDA_distTemps}, where the upper panels show the sampling for the different temperatures and the lower row shows the pair distance distribution.
From here, we can see how the pair distance distribution function evolves from a multimodal histogram at 100K to a less complex function as the temperature increases.
In terms of trajectory reconstruction, from Table~\ref{table:TempTrajLearn_MAEdist_normalized}, we can quickly see that when increasing the temperature of the reference data, the accuracy in the interatomic distances marginally decreases but the accuracy of the velocity field reconstruction considerably increases.
Furthermore, in the case of the Bi-LSTM model with $\Delta t=20$, the prediction accuracy increases roughly by a factor of 4 when training on data generated at 500K relative to the 100K case.

In a broader picture of the physical problem, these results tell us that the generated dynamics at higher temperatures are smoother even though the atomic speeds are higher. 
This is actually because the molecular system spends more time in anharmonic regions, which then generates the well-known frequency red shift of most of the vibrational frequencies~\cite{sGDML_Appl_jcp}.
In other words, increasing the system's temperature reduces the oscillation periods and generates larger oscillation amplitudes, thereby reducing the complexity of the learning problem.


\begin{figure*}[t]
\centering
\includegraphics[width=1.9\columnwidth]{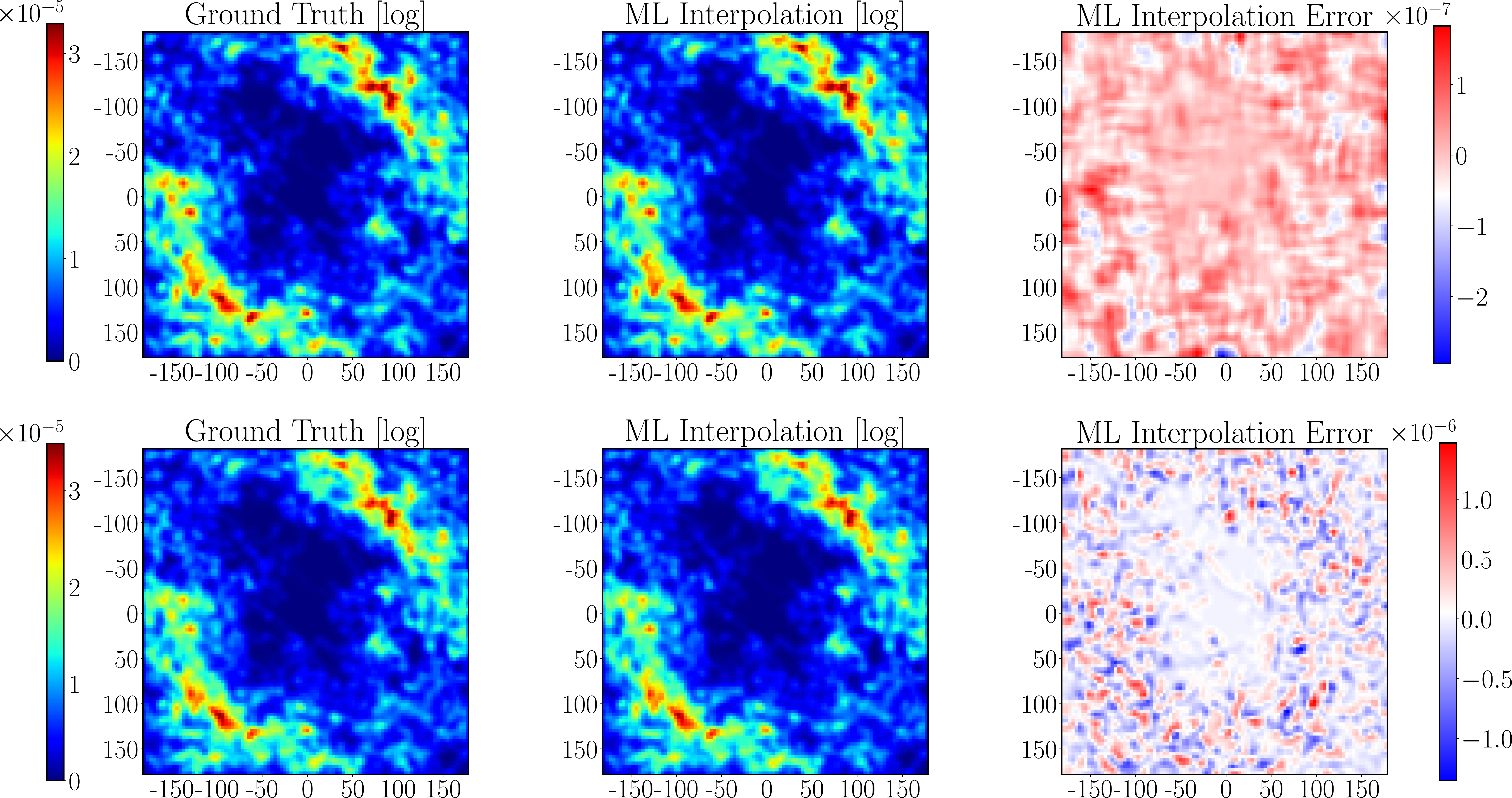}
\caption{Sampling generated from molecular dynamics using (right) the sGDML force field trained on density functional theory (MD17 database~\cite{gdml}) for the MDA, (middle) and its interpolation generated using a bi-directional LSTM model for $\Delta t=80$ fs. (Right) Difference between the reference and the interpolated trajectories.}
\label{fig:Malondialdehyde_DiffHist}
\end{figure*}

\begin{figure}[h!]
\centering
\includegraphics[width=1.0\columnwidth]{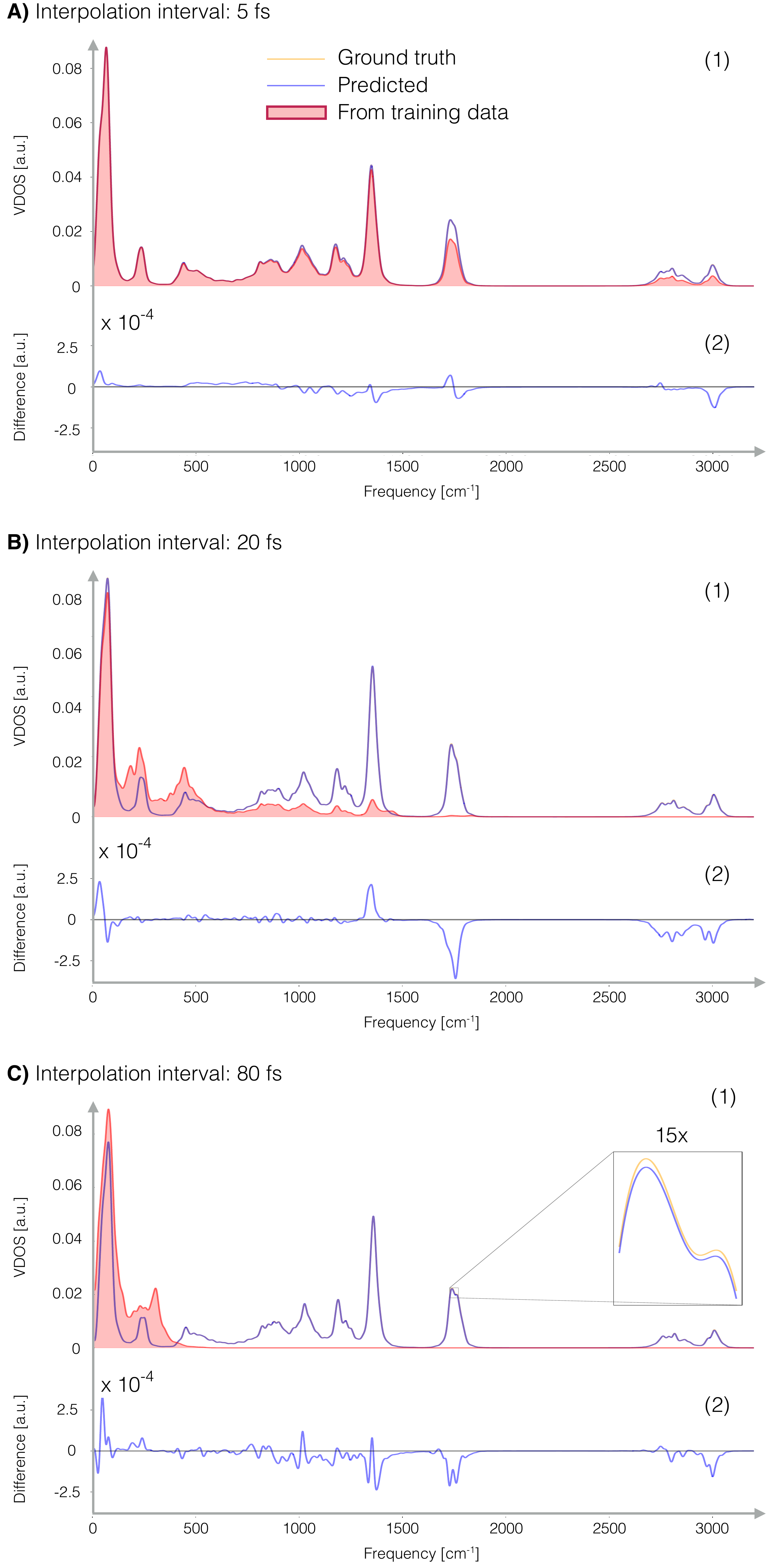}
\caption{Reconstruction of the vibrational density of states (VDOS) using the interpolation intervals: A) 5fs, B) 20 fs, and C) 80 fs. In panel (1) are the ground truth VDOS (yellow), the predicted VDOS (blue), and the VDOS obtained from the training data (shaded red). In Panel (2) is the differences between the ground truth and predicted VDOSs.}
\label{fig:VDOS_diffInter}
\end{figure}

\subsubsection{Free energy surface reconstruction}

To complement the results from the previous sections, here we correlate those results to the free energy surface (FES).
The FES can be estimated by integrating the probability distribution generated by the molecular trajectories, hence smoother trajectories in principle should lead to less complex FES as shown in Fig.~\ref{fig:HistkMDA_distTemps}.
Now, this actually has a beneficial connection to the learning problem, because given the previous analysis, we can accurately reconstruct the molecular FES from a reduced amount of data or sparse trajectories.
In order to corroborate such accuracy, in Fig.~\ref{fig:Malondialdehyde_DiffHist} we show a comparison between the reference and Bi-LSTM-predicted configurations' sampling for the keto-MDA molecule for its main degrees of freedom.
As mentioned before, the accuracy in trajectory reconstruction is $\sim10^{-4}$ \AA, meaning that the differences on the generated FES are barely noticeable, hence we highlight some differences by a letter \textbf{A} which is just an artefact coming from the binning procedure in the figure.

\subsubsection{Testing the limits of the methodology: Larger interpolation intervals}

In order to assess the accuracy of the bi-LSTM models as a function of the sampling frequency, in Fig.~\ref{fig:VDOS_diffInter} we show their different models with an interpolation capacity of 5, 20 and 80 fs, where the resolution of the training data is 1 fs.
This means that during training, the 80 fs model for example, was trained using time series sampled every 80 fs, given the initial and final conditions, the model will predict the intermediate 78 points in the series.
If we analyse the information encoded in a time series sampled every 80 fs, we can clearly see that the training data set only contain information regarding molecular oscillations with frequencies of $\sim$500 cm$^{-1}$ at most, and the rest of the spectrum is not explicitly embedded in the data.
The frequency spectrum shown to each model during the training process is highlighted in red in Fig.~\ref{fig:VDOS_diffInter}.
From this figure we obtain that in the three cases, the reconstructed vibrational spectrum is indistinguishable from the ground truth calculations (see bottom plot in each of the panels), despite the wide variability of the interpolation parameter.
Nevertheless, in the 80 fs case (inset in Fig.~\ref{fig:VDOS_diffInter}-C), we can appreciate that some slight deviations start to appear.

These results immediately bring up the question of how the model can still reconstruct very high frequencies when these are not explicitly given.
The answer lies on the memory cells of the LSTM architectures.
It doesn't matter that high frequencies are not shown to the model, since in each pass of the training process, the model is getting different configurations that sample the high frequency normal modes. 
Hence, by keeping in memory all that information, the model can infer the existence of all the vibrations in the system. 

From the results in this section and the previous one, we obtain that bi-LSTM models not only manage to accurately reconstruct the spacial components of the trajectories, but also the velocity field.
Thereby, ensuring their application to reconstruct the phase-apace dynamics as well as the physical properties such as free energies and vibrational spectra.

\section{Conclusions}

In this work we have introduced a series of bi-directional NN integrators (i.e. NODE, HNN, RNN and LSTM) to increase the resolution of molecular dynamics trajectories for a variety of molecular systems. 
%
An extensive validation process of these models on the MD17 dataset and different interpolation resolutions was presented (see Fig.~\ref{fig:performance}).
For here, we found that the Bi-LSTM model is the better performing method generating more stable results across all the extended-MD17 dataset and better accuracies, reaching errors as low as 10$^{-4}$ \AA.
These errors render the interpolated and ground truth trajectories indistinguishable.
The overall higher performance of Bi-LSTM networks is due to the robustness against noisy (thermostated) reference data, as well as their capacity to retain long time correlations' during the learning process.

Beyond the machine learning task of high dimensional trajectory interpolation, by varying the temperature of the reference data used for training the models, we obtained important physical insights regarding the dynamical behaviour of molecular systems.
There is evidence in the literature that learning molecular force fields gets more complicated as the temperature of the training data increases.
Contrasting this behaviour, here we have observed that trajectory interpolation becomes easier as the temperature increases.
The origin of such different temperature-dependent learning behaviour is the fact that the trajectories generated by molecular dynamics simulations effectively move on the Helmholtz free energy surface due to entropic contributions.
In other words, finite temperature trajectories sample the FES instead of the underlying PES.
Furthermore, the FES gets smoother as the temperature increases, meaning that the FES' landscape progressively deviates from the PES as the temperature increases.
Consequently, reconstructing the underlying PES from MD datasets becomes harder as the temperature increases, while this benefits FES learning.
Additionally, we found that the remarkable property of Bi-LSTMs of being able to learn the full spectrum of molecular vibrational frequencies (harmonic and anharmonic) even though this information is not explicitly shown during the training process. This capability is due to their memory cells. 

From the results obtained in this work, we have evinced the great learning capacity of Bi-LSTMs on the reconstruction of realistic high dimensional molecular behaviour.
This opens up a new set of applications for the family of recurrent neural networks on post-processing molecular dynamics results.
Furthermore, the results here presented can be used to set the stage for robust extrapolation techniques, either by using it as a data enhancement method for forecasting or as the basis for numerical propagator.

\section*{Data Availability Statement}
The training data  and code base used in this study are available from the authors upon reasonable request.

\newpage
\bibliography{references}

\begin{thebibliography}{42}%
\makeatletter
\providecommand \@ifxundefined [1]{%
 \@ifx{#1\undefined}
}%
\providecommand \@ifnum [1]{%
 \ifnum #1\expandafter \@firstoftwo
 \else \expandafter \@secondoftwo
 \fi
}%
\providecommand \@ifx [1]{%
 \ifx #1\expandafter \@firstoftwo
 \else \expandafter \@secondoftwo
 \fi
}%
\providecommand \natexlab [1]{#1}%
\providecommand \enquote  [1]{``#1''}%
\providecommand \bibnamefont  [1]{#1}%
\providecommand \bibfnamefont [1]{#1}%
\providecommand \citenamefont [1]{#1}%
\providecommand \href@noop [0]{\@secondoftwo}%
\providecommand \href [0]{\begingroup \@sanitize@url \@href}%
\providecommand \@href[1]{\@@startlink{#1}\@@href}%
\providecommand \@@href[1]{\endgroup#1\@@endlink}%
\providecommand \@sanitize@url [0]{\catcode `\\12\catcode `\$12\catcode
  `\&12\catcode `\#12\catcode `\^12\catcode `\_12\catcode `\%12\relax}%
\providecommand \@@startlink[1]{}%
\providecommand \@@endlink[0]{}%
\providecommand \url  [0]{\begingroup\@sanitize@url \@url }%
\providecommand \@url [1]{\endgroup\@href {#1}{\urlprefix }}%
\providecommand \urlprefix  [0]{URL }%
\providecommand \Eprint [0]{\href }%
\providecommand \doibase [0]{http://dx.doi.org/}%
\providecommand \selectlanguage [0]{\@gobble}%
\providecommand \bibinfo  [0]{\@secondoftwo}%
\providecommand \bibfield  [0]{\@secondoftwo}%
\providecommand \translation [1]{[#1]}%
\providecommand \BibitemOpen [0]{}%
\providecommand \bibitemStop [0]{}%
\providecommand \bibitemNoStop [0]{.\EOS\space}%
\providecommand \EOS [0]{\spacefactor3000\relax}%
\providecommand \BibitemShut  [1]{\csname bibitem#1\endcsname}%
\let\auto@bib@innerbib\@empty
\bibitem [{\citenamefont {Tuckerman}(2010)}]{tuckerman2010statistical}%
  \BibitemOpen
  \bibfield  {author} {\bibinfo {author} {\bibfnamefont {M.}~\bibnamefont
  {Tuckerman}},\ }\href {\doibase 10.1002/ange.201105752} {\emph {\bibinfo
  {title} {Statistical Mechanics: Theory and Molecular Simulation}}}\ (\bibinfo
   {publisher} {Oxford University Press},\ \bibinfo {year} {2010})\BibitemShut
  {NoStop}%
\bibitem [{\citenamefont {No{\'e}}\ \emph {et~al.}(2020)\citenamefont
  {No{\'e}}, \citenamefont {Tkatchenko}, \citenamefont {M{\"u}ller},\ and\
  \citenamefont {Clementi}}]{noe2020machine}%
  \BibitemOpen
  \bibfield  {author} {\bibinfo {author} {\bibfnamefont {F.}~\bibnamefont
  {No{\'e}}}, \bibinfo {author} {\bibfnamefont {A.}~\bibnamefont {Tkatchenko}},
  \bibinfo {author} {\bibfnamefont {K.-R.}\ \bibnamefont {M{\"u}ller}}, \ and\
  \bibinfo {author} {\bibfnamefont {C.}~\bibnamefont {Clementi}},\ }\href@noop
  {} {\bibfield  {journal} {\bibinfo  {journal} {Annual review of physical
  chemistry}\ }\textbf {\bibinfo {volume} {71}},\ \bibinfo {pages} {361}
  (\bibinfo {year} {2020})}\BibitemShut {NoStop}%
\bibitem [{\citenamefont {Sadeghi}\ and\ \citenamefont
  {No{\'e}}(2021)}]{Noe2021_JPCLett}%
  \BibitemOpen
  \bibfield  {author} {\bibinfo {author} {\bibfnamefont {M.}~\bibnamefont
  {Sadeghi}}\ and\ \bibinfo {author} {\bibfnamefont {F.}~\bibnamefont
  {No{\'e}}},\ }\href {\doibase 10.1021/acs.jpclett.1c02954} {\bibfield
  {journal} {\bibinfo  {journal} {J. Phys. Chem. Lett.}\ }\textbf {\bibinfo
  {volume} {12}},\ \bibinfo {pages} {10497} (\bibinfo {year}
  {2021})}\BibitemShut {NoStop}%
\bibitem [{\citenamefont {Behler}(2017)}]{Behler.ACIE56.2017}%
  \BibitemOpen
  \bibfield  {author} {\bibinfo {author} {\bibfnamefont {J.}~\bibnamefont
  {Behler}},\ }\href {\doibase 10.1002/anie.201703114} {\bibfield  {journal}
  {\bibinfo  {journal} {Angew. Chem. Int. Ed.}\ }\textbf {\bibinfo {volume}
  {56}},\ \bibinfo {pages} {12828} (\bibinfo {year} {2017})}\BibitemShut
  {NoStop}%
\bibitem [{\citenamefont {Sch\"{u}tt}\ \emph {et~al.}(2017)\citenamefont
  {Sch\"{u}tt}, \citenamefont {Kindermans}, \citenamefont {Sauceda},
  \citenamefont {Chmiela}, \citenamefont {Tkatchenko},\ and\ \citenamefont
  {M\"{u}ller}}]{SchNetNIPS2017}%
  \BibitemOpen
  \bibfield  {author} {\bibinfo {author} {\bibfnamefont {K.~T.}\ \bibnamefont
  {Sch\"{u}tt}}, \bibinfo {author} {\bibfnamefont {P.-J.}\ \bibnamefont
  {Kindermans}}, \bibinfo {author} {\bibfnamefont {H.~E.}\ \bibnamefont
  {Sauceda}}, \bibinfo {author} {\bibfnamefont {S.}~\bibnamefont {Chmiela}},
  \bibinfo {author} {\bibfnamefont {A.}~\bibnamefont {Tkatchenko}}, \ and\
  \bibinfo {author} {\bibfnamefont {K.-R.}\ \bibnamefont {M\"{u}ller}},\ }in\
  \href
  {http://papers.nips.cc/paper/6700-schnet-a-continuous-filter-convolutional-neural-network-for-modeling-quantum-interactions.pdf}
  {\emph {\bibinfo {booktitle} {Advances in Neural Information Processing
  Systems 30}}}\ (\bibinfo {year} {2017})\ pp.\ \bibinfo {pages}
  {991--1001}\BibitemShut {NoStop}%
\bibitem [{\citenamefont {Sch\"utt}\ \emph {et~al.}(2018)\citenamefont
  {Sch\"utt}, \citenamefont {Sauceda}, \citenamefont {Kindermans},
  \citenamefont {Tkatchenko},\ and\ \citenamefont {M\"uller}}]{SchNet2018}%
  \BibitemOpen
  \bibfield  {author} {\bibinfo {author} {\bibfnamefont {K.~T.}\ \bibnamefont
  {Sch\"utt}}, \bibinfo {author} {\bibfnamefont {H.~E.}\ \bibnamefont
  {Sauceda}}, \bibinfo {author} {\bibfnamefont {P.-J.}\ \bibnamefont
  {Kindermans}}, \bibinfo {author} {\bibfnamefont {A.}~\bibnamefont
  {Tkatchenko}}, \ and\ \bibinfo {author} {\bibfnamefont {K.-R.}\ \bibnamefont
  {M\"uller}},\ }\href {\doibase 10.1063/1.5019779} {\bibfield  {journal}
  {\bibinfo  {journal} {J. Chem. Phys.}\ }\textbf {\bibinfo {volume} {148}},\
  \bibinfo {pages} {241722} (\bibinfo {year} {2018})}\BibitemShut {NoStop}%
\bibitem [{\citenamefont {Han}\ \emph {et~al.}(2018)\citenamefont {Han},
  \citenamefont {Zhang}, \citenamefont {Car},\ and\ \citenamefont
  {E}}]{Han_DeeP_2018}%
  \BibitemOpen
  \bibfield  {author} {\bibinfo {author} {\bibfnamefont {J.}~\bibnamefont
  {Han}}, \bibinfo {author} {\bibfnamefont {L.}~\bibnamefont {Zhang}}, \bibinfo
  {author} {\bibfnamefont {R.}~\bibnamefont {Car}}, \ and\ \bibinfo {author}
  {\bibfnamefont {W.}~\bibnamefont {E}},\ }\href {\doibase
  https://doi.org/10.4208/cicp.OA-2017-0213} {\bibfield  {journal} {\bibinfo
  {journal} {Commun. Comput. Phys.}\ }\textbf {\bibinfo {volume} {23}},\
  \bibinfo {pages} {629} (\bibinfo {year} {2018})}\BibitemShut {NoStop}%
\bibitem [{\citenamefont {Chmiela}\ \emph {et~al.}(2017)\citenamefont
  {Chmiela}, \citenamefont {Tkatchenko}, \citenamefont {Sauceda}, \citenamefont
  {Poltavsky}, \citenamefont {Sch{\"u}tt},\ and\ \citenamefont
  {M{\"u}ller}}]{gdml}%
  \BibitemOpen
  \bibfield  {author} {\bibinfo {author} {\bibfnamefont {S.}~\bibnamefont
  {Chmiela}}, \bibinfo {author} {\bibfnamefont {A.}~\bibnamefont {Tkatchenko}},
  \bibinfo {author} {\bibfnamefont {H.~E.}\ \bibnamefont {Sauceda}}, \bibinfo
  {author} {\bibfnamefont {I.}~\bibnamefont {Poltavsky}}, \bibinfo {author}
  {\bibfnamefont {K.~T.}\ \bibnamefont {Sch{\"u}tt}}, \ and\ \bibinfo {author}
  {\bibfnamefont {K.-R.}\ \bibnamefont {M{\"u}ller}},\ }\href {\doibase
  10.1126/sciadv.1603015} {\bibfield  {journal} {\bibinfo  {journal} {Sci.
  Adv.}\ }\textbf {\bibinfo {volume} {3}},\ \bibinfo {pages} {e1603015}
  (\bibinfo {year} {2017})}\BibitemShut {NoStop}%
\bibitem [{\citenamefont {Chmiela}\ \emph {et~al.}(2018)\citenamefont
  {Chmiela}, \citenamefont {Sauceda}, \citenamefont {M{\"u}ller},\ and\
  \citenamefont {Tkatchenko}}]{sgdml}%
  \BibitemOpen
  \bibfield  {author} {\bibinfo {author} {\bibfnamefont {S.}~\bibnamefont
  {Chmiela}}, \bibinfo {author} {\bibfnamefont {H.~E.}\ \bibnamefont
  {Sauceda}}, \bibinfo {author} {\bibfnamefont {K.-R.}\ \bibnamefont
  {M{\"u}ller}}, \ and\ \bibinfo {author} {\bibfnamefont {A.}~\bibnamefont
  {Tkatchenko}},\ }\href {\doibase 10.1038/s41467-018-06169-2} {\bibfield
  {journal} {\bibinfo  {journal} {Nat. Commun.}\ }\textbf {\bibinfo {volume}
  {9}},\ \bibinfo {pages} {3887} (\bibinfo {year} {2018})}\BibitemShut
  {NoStop}%
\bibitem [{\citenamefont {Novikov}\ \emph {et~al.}(2021)\citenamefont
  {Novikov}, \citenamefont {Gubaev}, \citenamefont {Podryabinkin},\ and\
  \citenamefont {Shapeev}}]{Shapeev_Moment_MLSciT2021}%
  \BibitemOpen
  \bibfield  {author} {\bibinfo {author} {\bibfnamefont {I.~S.}\ \bibnamefont
  {Novikov}}, \bibinfo {author} {\bibfnamefont {K.}~\bibnamefont {Gubaev}},
  \bibinfo {author} {\bibfnamefont {E.~V.}\ \bibnamefont {Podryabinkin}}, \
  and\ \bibinfo {author} {\bibfnamefont {A.~V.}\ \bibnamefont {Shapeev}},\
  }\href {\doibase 10.1088/2632-2153/abc9fe} {\bibfield  {journal} {\bibinfo
  {journal} {Mach. Learn.: Sci. Technol.}\ }\textbf {\bibinfo {volume} {2}},\
  \bibinfo {pages} {025002} (\bibinfo {year} {2021})}\BibitemShut {NoStop}%
\bibitem [{\citenamefont {Bart\'ok}\ \emph {et~al.}(2010)\citenamefont
  {Bart\'ok}, \citenamefont {Payne}, \citenamefont {Kondor},\ and\
  \citenamefont {Cs\'anyi}}]{Bartok_GAP_PRL2010}%
  \BibitemOpen
  \bibfield  {author} {\bibinfo {author} {\bibfnamefont {A.~P.}\ \bibnamefont
  {Bart\'ok}}, \bibinfo {author} {\bibfnamefont {M.~C.}\ \bibnamefont {Payne}},
  \bibinfo {author} {\bibfnamefont {R.}~\bibnamefont {Kondor}}, \ and\ \bibinfo
  {author} {\bibfnamefont {G.}~\bibnamefont {Cs\'anyi}},\ }\href {\doibase
  10.1103/PhysRevLett.104.136403} {\bibfield  {journal} {\bibinfo  {journal}
  {Phys. Rev. Lett.}\ }\textbf {\bibinfo {volume} {104}},\ \bibinfo {pages}
  {136403} (\bibinfo {year} {2010})}\BibitemShut {NoStop}%
\bibitem [{\citenamefont {Zeni}\ \emph {et~al.}(2018)\citenamefont {Zeni},
  \citenamefont {Rossi}, \citenamefont {Glielmo}, \citenamefont {Fekete},
  \citenamefont {Gaston}, \citenamefont {Baletto},\ and\ \citenamefont
  {De~Vita}}]{Zeni_De_Vita2018}%
  \BibitemOpen
  \bibfield  {author} {\bibinfo {author} {\bibfnamefont {C.}~\bibnamefont
  {Zeni}}, \bibinfo {author} {\bibfnamefont {K.}~\bibnamefont {Rossi}},
  \bibinfo {author} {\bibfnamefont {A.}~\bibnamefont {Glielmo}}, \bibinfo
  {author} {\bibfnamefont {A.}~\bibnamefont {Fekete}}, \bibinfo {author}
  {\bibfnamefont {N.}~\bibnamefont {Gaston}}, \bibinfo {author} {\bibfnamefont
  {F.}~\bibnamefont {Baletto}}, \ and\ \bibinfo {author} {\bibfnamefont
  {A.}~\bibnamefont {De~Vita}},\ }\href {\doibase 10.1063/1.5024558} {\bibfield
   {journal} {\bibinfo  {journal} {J. Chem. Phys.}\ }\textbf {\bibinfo {volume}
  {148}},\ \bibinfo {pages} {241739} (\bibinfo {year} {2018})}\BibitemShut
  {NoStop}%
\bibitem [{\citenamefont {Glielmo}\ \emph {et~al.}(2018)\citenamefont
  {Glielmo}, \citenamefont {Zeni},\ and\ \citenamefont
  {De~Vita}}]{Glielmo.PRB97.2018}%
  \BibitemOpen
  \bibfield  {author} {\bibinfo {author} {\bibfnamefont {A.}~\bibnamefont
  {Glielmo}}, \bibinfo {author} {\bibfnamefont {C.}~\bibnamefont {Zeni}}, \
  and\ \bibinfo {author} {\bibfnamefont {A.}~\bibnamefont {De~Vita}},\ }\href
  {\doibase 10.1103/PhysRevB.97.184307} {\bibfield  {journal} {\bibinfo
  {journal} {Phys. Rev. B}\ }\textbf {\bibinfo {volume} {97}},\ \bibinfo
  {pages} {184307} (\bibinfo {year} {2018})}\BibitemShut {NoStop}%
\bibitem [{\citenamefont {Unke}\ \emph
  {et~al.}(2021{\natexlab{a}})\citenamefont {Unke}, \citenamefont {Chmiela},
  \citenamefont {Sauceda}, \citenamefont {Gastegger}, \citenamefont
  {Poltavsky}, \citenamefont {Sch{\"u}tt}, \citenamefont {Tkatchenko},\ and\
  \citenamefont {M{\"u}ller}}]{unke2020machine}%
  \BibitemOpen
  \bibfield  {author} {\bibinfo {author} {\bibfnamefont {O.~T.}\ \bibnamefont
  {Unke}}, \bibinfo {author} {\bibfnamefont {S.}~\bibnamefont {Chmiela}},
  \bibinfo {author} {\bibfnamefont {H.~E.}\ \bibnamefont {Sauceda}}, \bibinfo
  {author} {\bibfnamefont {M.}~\bibnamefont {Gastegger}}, \bibinfo {author}
  {\bibfnamefont {I.}~\bibnamefont {Poltavsky}}, \bibinfo {author}
  {\bibfnamefont {K.~T.}\ \bibnamefont {Sch{\"u}tt}}, \bibinfo {author}
  {\bibfnamefont {A.}~\bibnamefont {Tkatchenko}}, \ and\ \bibinfo {author}
  {\bibfnamefont {K.-R.}\ \bibnamefont {M{\"u}ller}},\ }\href {\doibase
  https://doi.org/10.1021/acs.chemrev.0c01111} {\bibfield  {journal} {\bibinfo
  {journal} {Chem. Rev.}\ }\textbf {\bibinfo {volume} {121}},\ \bibinfo {pages}
  {10142} (\bibinfo {year} {2021}{\natexlab{a}})}\BibitemShut {NoStop}%
\bibitem [{\citenamefont {von Lilienfeld}\ \emph {et~al.}(2020)\citenamefont
  {von Lilienfeld}, \citenamefont {M{\"u}ller},\ and\ \citenamefont
  {Tkatchenko}}]{von2020exploring}%
  \BibitemOpen
  \bibfield  {author} {\bibinfo {author} {\bibfnamefont {O.~A.}\ \bibnamefont
  {von Lilienfeld}}, \bibinfo {author} {\bibfnamefont {K.-R.}\ \bibnamefont
  {M{\"u}ller}}, \ and\ \bibinfo {author} {\bibfnamefont {A.}~\bibnamefont
  {Tkatchenko}},\ }\href {\doibase 10.1038/s41570-020-0189-9} {\bibfield
  {journal} {\bibinfo  {journal} {Nat. Rev. Chem.}\ }\textbf {\bibinfo {volume}
  {4}},\ \bibinfo {pages} {347–} (\bibinfo {year} {2020})}\BibitemShut
  {NoStop}%
\bibitem [{\citenamefont {Tkatchenko}(2020)}]{Tkatchenko2020NatCommun}%
  \BibitemOpen
  \bibfield  {author} {\bibinfo {author} {\bibfnamefont {A.}~\bibnamefont
  {Tkatchenko}},\ }\href {\doibase https://doi.org/10.1038/s41467-020-17844-8}
  {\bibfield  {journal} {\bibinfo  {journal} {Nat. Commun.}\ }\textbf {\bibinfo
  {volume} {11}},\ \bibinfo {pages} {4125} (\bibinfo {year}
  {2020})}\BibitemShut {NoStop}%
\bibitem [{\citenamefont {Musil}\ \emph {et~al.}(2021)\citenamefont {Musil},
  \citenamefont {Grisafi}, \citenamefont {Bart{\'o}k}, \citenamefont {Ortner},
  \citenamefont {Cs{\'a}nyi},\ and\ \citenamefont
  {Ceriotti}}]{musil2021physics}%
  \BibitemOpen
  \bibfield  {author} {\bibinfo {author} {\bibfnamefont {F.}~\bibnamefont
  {Musil}}, \bibinfo {author} {\bibfnamefont {A.}~\bibnamefont {Grisafi}},
  \bibinfo {author} {\bibfnamefont {A.~P.}\ \bibnamefont {Bart{\'o}k}},
  \bibinfo {author} {\bibfnamefont {C.}~\bibnamefont {Ortner}}, \bibinfo
  {author} {\bibfnamefont {G.}~\bibnamefont {Cs{\'a}nyi}}, \ and\ \bibinfo
  {author} {\bibfnamefont {M.}~\bibnamefont {Ceriotti}},\ }\href {\doibase
  https://doi.org/10.1021/acs.chemrev.1c00021} {\bibfield  {journal} {\bibinfo
  {journal} {Chem. Rev.}\ }\textbf {\bibinfo {volume} {121}},\ \bibinfo {pages}
  {9759–9815} (\bibinfo {year} {2021})}\BibitemShut {NoStop}%
\bibitem [{\citenamefont {Keith}\ \emph {et~al.}(2021)\citenamefont {Keith},
  \citenamefont {Vassilev-Galindo}, \citenamefont {Cheng}, \citenamefont
  {Chmiela}, \citenamefont {Gastegger}, \citenamefont {M{\"u}ller},\ and\
  \citenamefont {Tkatchenko}}]{keith2021combining}%
  \BibitemOpen
  \bibfield  {author} {\bibinfo {author} {\bibfnamefont {J.~A.}\ \bibnamefont
  {Keith}}, \bibinfo {author} {\bibfnamefont {V.}~\bibnamefont
  {Vassilev-Galindo}}, \bibinfo {author} {\bibfnamefont {B.}~\bibnamefont
  {Cheng}}, \bibinfo {author} {\bibfnamefont {S.}~\bibnamefont {Chmiela}},
  \bibinfo {author} {\bibfnamefont {M.}~\bibnamefont {Gastegger}}, \bibinfo
  {author} {\bibfnamefont {K.-R.}\ \bibnamefont {M{\"u}ller}}, \ and\ \bibinfo
  {author} {\bibfnamefont {A.}~\bibnamefont {Tkatchenko}},\ }\href {\doibase
  10.1021/acs.chemrev.1c00107} {\bibfield  {journal} {\bibinfo  {journal}
  {Chem. Rev.}\ }\textbf {\bibinfo {volume} {121}},\ \bibinfo {pages}
  {9816–9872} (\bibinfo {year} {2021})}\BibitemShut {NoStop}%
\bibitem [{\citenamefont {Unke}\ \emph
  {et~al.}(2021{\natexlab{b}})\citenamefont {Unke}, \citenamefont {Chmiela},
  \citenamefont {Gastegger}, \citenamefont {Sch\"utt}, \citenamefont
  {Sauceda},\ and\ \citenamefont {M\"uller}}]{unke2021spookynet}%
  \BibitemOpen
  \bibfield  {author} {\bibinfo {author} {\bibfnamefont {O.~T.}\ \bibnamefont
  {Unke}}, \bibinfo {author} {\bibfnamefont {S.}~\bibnamefont {Chmiela}},
  \bibinfo {author} {\bibfnamefont {M.}~\bibnamefont {Gastegger}}, \bibinfo
  {author} {\bibfnamefont {K.~T.}\ \bibnamefont {Sch\"utt}}, \bibinfo {author}
  {\bibfnamefont {H.~E.}\ \bibnamefont {Sauceda}}, \ and\ \bibinfo {author}
  {\bibfnamefont {K.-R.}\ \bibnamefont {M\"uller}},\ }\href {\doibase
  10.1038/s41467-021-27504-0} {\bibfield  {journal} {\bibinfo  {journal} {Nat.
  Commun.}\ ,\ \bibinfo {pages} {7273}} (\bibinfo {year}
  {2021}{\natexlab{b}})}\BibitemShut {NoStop}%
\bibitem [{\citenamefont {Sauceda}\ \emph
  {et~al.}(2021{\natexlab{a}})\citenamefont {Sauceda}, \citenamefont
  {Gálvez-González}, \citenamefont {Chmiela}, \citenamefont {Paz-Borbón},
  \citenamefont {Müller},\ and\ \citenamefont {Tkatchenko}}]{Sauceda_bigdml}%
  \BibitemOpen
  \bibfield  {author} {\bibinfo {author} {\bibfnamefont {H.~E.}\ \bibnamefont
  {Sauceda}}, \bibinfo {author} {\bibfnamefont {L.~E.}\ \bibnamefont
  {Gálvez-González}}, \bibinfo {author} {\bibfnamefont {S.}~\bibnamefont
  {Chmiela}}, \bibinfo {author} {\bibfnamefont {L.~O.}\ \bibnamefont
  {Paz-Borbón}}, \bibinfo {author} {\bibfnamefont {K.-R.}\ \bibnamefont
  {Müller}}, \ and\ \bibinfo {author} {\bibfnamefont {A.}~\bibnamefont
  {Tkatchenko}},\ }\href@noop {} {\  (\bibinfo {year} {2021}{\natexlab{a}})},\
  \Eprint {http://arxiv.org/abs/arXiv:2106.04229} {arXiv:2106.04229}
  \BibitemShut {NoStop}%
\bibitem [{\citenamefont {Li}\ \emph {et~al.}(2021)\citenamefont {Li},
  \citenamefont {Meidani}, \citenamefont {Yadav},\ and\ \citenamefont
  {Farimani}}]{li2021graphNNFF}%
  \BibitemOpen
  \bibfield  {author} {\bibinfo {author} {\bibfnamefont {Z.}~\bibnamefont
  {Li}}, \bibinfo {author} {\bibfnamefont {K.}~\bibnamefont {Meidani}},
  \bibinfo {author} {\bibfnamefont {P.}~\bibnamefont {Yadav}}, \ and\ \bibinfo
  {author} {\bibfnamefont {A.~B.}\ \bibnamefont {Farimani}},\ }\href@noop {}
  {\enquote {\bibinfo {title} {Graph neural networks accelerated molecular
  dynamics},}\ } (\bibinfo {year} {2021}),\ \Eprint
  {http://arxiv.org/abs/2112.03383} {arXiv:2112.03383 [cs.LG]} \BibitemShut
  {NoStop}%
\bibitem [{\citenamefont {Sauceda}\ \emph
  {et~al.}(2021{\natexlab{b}})\citenamefont {Sauceda}, \citenamefont
  {Vassilev-Galindo}, \citenamefont {Chmiela}, \citenamefont {M{\"u}ller},\
  and\ \citenamefont {Tkatchenko}}]{sauceda_NQE_NatComm2021}%
  \BibitemOpen
  \bibfield  {author} {\bibinfo {author} {\bibfnamefont {H.~E.}\ \bibnamefont
  {Sauceda}}, \bibinfo {author} {\bibfnamefont {V.}~\bibnamefont
  {Vassilev-Galindo}}, \bibinfo {author} {\bibfnamefont {S.}~\bibnamefont
  {Chmiela}}, \bibinfo {author} {\bibfnamefont {K.-R.}\ \bibnamefont
  {M{\"u}ller}}, \ and\ \bibinfo {author} {\bibfnamefont {A.}~\bibnamefont
  {Tkatchenko}},\ }\href {\doibase 10.1038/s41467-020-20212-1} {\bibfield
  {journal} {\bibinfo  {journal} {Nat. Commun.}\ }\textbf {\bibinfo {volume}
  {12}},\ \bibinfo {pages} {442} (\bibinfo {year}
  {2021}{\natexlab{b}})}\BibitemShut {NoStop}%
\bibitem [{\citenamefont {Tsai}\ \emph {et~al.}(2020)\citenamefont {Tsai},
  \citenamefont {Kuo},\ and\ \citenamefont {Tiwary}}]{tsai2020LMMD}%
  \BibitemOpen
  \bibfield  {author} {\bibinfo {author} {\bibfnamefont {S.-T.}\ \bibnamefont
  {Tsai}}, \bibinfo {author} {\bibfnamefont {E.-J.}\ \bibnamefont {Kuo}}, \
  and\ \bibinfo {author} {\bibfnamefont {P.}~\bibnamefont {Tiwary}},\ }\href
  {\doibase 10.1038/s41467-020-18959-8} {\bibfield  {journal} {\bibinfo
  {journal} {Nat. Commun.}\ }\textbf {\bibinfo {volume} {11}},\ \bibinfo
  {pages} {5115} (\bibinfo {year} {2020})}\BibitemShut {NoStop}%
\bibitem [{\citenamefont {Thaler}\ and\ \citenamefont
  {Zavadlav}(2021)}]{thaler_traj_2021}%
  \BibitemOpen
  \bibfield  {author} {\bibinfo {author} {\bibfnamefont {S.}~\bibnamefont
  {Thaler}}\ and\ \bibinfo {author} {\bibfnamefont {J.}~\bibnamefont
  {Zavadlav}},\ }\href@noop {} {\  (\bibinfo {year} {2021})},\ \Eprint
  {http://arxiv.org/abs/2106.01138} {arXiv:2106.01138} \BibitemShut {NoStop}%
\bibitem [{\citenamefont {Vlachas}\ \emph
  {et~al.}(2021{\natexlab{a}})\citenamefont {Vlachas}, \citenamefont
  {Arampatzis}, \citenamefont {Uhler},\ and\ \citenamefont
  {Koumoutsakos}}]{vlachas2021multiscale}%
  \BibitemOpen
  \bibfield  {author} {\bibinfo {author} {\bibfnamefont {P.~R.}\ \bibnamefont
  {Vlachas}}, \bibinfo {author} {\bibfnamefont {G.}~\bibnamefont {Arampatzis}},
  \bibinfo {author} {\bibfnamefont {C.}~\bibnamefont {Uhler}}, \ and\ \bibinfo
  {author} {\bibfnamefont {P.}~\bibnamefont {Koumoutsakos}},\ }\href@noop {}
  {\enquote {\bibinfo {title} {Multiscale simulations of complex systems by
  learning their effective dynamics},}\ } (\bibinfo {year}
  {2021}{\natexlab{a}}),\ \Eprint {http://arxiv.org/abs/2006.13431}
  {arXiv:2006.13431 [physics.comp-ph]} \BibitemShut {NoStop}%
\bibitem [{\citenamefont {Vlachas}\ \emph
  {et~al.}(2021{\natexlab{b}})\citenamefont {Vlachas}, \citenamefont
  {Zavadlav}, \citenamefont {Praprotnik},\ and\ \citenamefont
  {Koumoutsakos}}]{vlachas2021accelerated}%
  \BibitemOpen
  \bibfield  {author} {\bibinfo {author} {\bibfnamefont {P.~R.}\ \bibnamefont
  {Vlachas}}, \bibinfo {author} {\bibfnamefont {J.}~\bibnamefont {Zavadlav}},
  \bibinfo {author} {\bibfnamefont {M.}~\bibnamefont {Praprotnik}}, \ and\
  \bibinfo {author} {\bibfnamefont {P.}~\bibnamefont {Koumoutsakos}},\
  }\href@noop {} {\enquote {\bibinfo {title} {Accelerated simulations of
  molecular systems through learning of their effective dynamics},}\ }
  (\bibinfo {year} {2021}{\natexlab{b}}),\ \Eprint
  {http://arxiv.org/abs/2102.08810} {arXiv:2102.08810 [physics.comp-ph]}
  \BibitemShut {NoStop}%
\bibitem [{\citenamefont {Kadupitiya}\ \emph {et~al.}(2020)\citenamefont
  {Kadupitiya}, \citenamefont {Fox},\ and\ \citenamefont
  {Jadhao}}]{kadupitiya2020simulating}%
  \BibitemOpen
  \bibfield  {author} {\bibinfo {author} {\bibfnamefont {J.}~\bibnamefont
  {Kadupitiya}}, \bibinfo {author} {\bibfnamefont {G.~C.}\ \bibnamefont {Fox}},
  \ and\ \bibinfo {author} {\bibfnamefont {V.}~\bibnamefont {Jadhao}},\
  }\href@noop {} {\  (\bibinfo {year} {2020})},\ \Eprint
  {http://arxiv.org/abs/2004.06493} {arXiv:2004.06493} \BibitemShut {NoStop}%
\bibitem [{\citenamefont {Chen}\ \emph {et~al.}(2018)\citenamefont {Chen},
  \citenamefont {Rubanova}, \citenamefont {Bettencourt},\ and\ \citenamefont
  {Duvenaud}}]{chen2018neural}%
  \BibitemOpen
  \bibfield  {author} {\bibinfo {author} {\bibfnamefont {R.~T.~Q.}\
  \bibnamefont {Chen}}, \bibinfo {author} {\bibfnamefont {Y.}~\bibnamefont
  {Rubanova}}, \bibinfo {author} {\bibfnamefont {J.}~\bibnamefont
  {Bettencourt}}, \ and\ \bibinfo {author} {\bibfnamefont {D.~K.}\ \bibnamefont
  {Duvenaud}},\ }in\ \href
  {https://proceedings.neurips.cc/paper/2018/file/69386f6bb1dfed68692a24c8686939b9-Paper.pdf}
  {\emph {\bibinfo {booktitle} {Advances in Neural Information Processing
  Systems}}},\ Vol.~\bibinfo {volume} {31}\ (\bibinfo  {publisher} {Curran
  Associates, Inc.},\ \bibinfo {year} {2018})\BibitemShut {NoStop}%
\bibitem [{\citenamefont {Greydanus}\ \emph {et~al.}(2019)\citenamefont
  {Greydanus}, \citenamefont {Dzamba},\ and\ \citenamefont
  {Yosinski}}]{greydanus2019hamiltonian}%
  \BibitemOpen
  \bibfield  {author} {\bibinfo {author} {\bibfnamefont {S.}~\bibnamefont
  {Greydanus}}, \bibinfo {author} {\bibfnamefont {M.}~\bibnamefont {Dzamba}}, \
  and\ \bibinfo {author} {\bibfnamefont {J.}~\bibnamefont {Yosinski}},\ }in\
  \href {\doibase arXiv:1909.04240} {\emph {\bibinfo {booktitle} {Adv. Neural
  Inf. Process. Sys.}}}\ (\bibinfo {year} {2019})\ pp.\ \bibinfo {pages}
  {15379--15389}\BibitemShut {NoStop}%
\bibitem [{\citenamefont {Cranmer}\ \emph {et~al.}(2020)\citenamefont
  {Cranmer}, \citenamefont {Greydanus}, \citenamefont {Hoyer}, \citenamefont
  {Battaglia}, \citenamefont {Spergel},\ and\ \citenamefont
  {Ho}}]{cranmer2020lagrangian}%
  \BibitemOpen
  \bibfield  {author} {\bibinfo {author} {\bibfnamefont {M.}~\bibnamefont
  {Cranmer}}, \bibinfo {author} {\bibfnamefont {S.}~\bibnamefont {Greydanus}},
  \bibinfo {author} {\bibfnamefont {S.}~\bibnamefont {Hoyer}}, \bibinfo
  {author} {\bibfnamefont {P.}~\bibnamefont {Battaglia}}, \bibinfo {author}
  {\bibfnamefont {D.}~\bibnamefont {Spergel}}, \ and\ \bibinfo {author}
  {\bibfnamefont {S.}~\bibnamefont {Ho}},\ }in\ \href {\doibase
  arXiv:2003.04630} {\emph {\bibinfo {booktitle} {Adv. Int. Conf. Learn.
  Repr.}}}\ (\bibinfo {year} {2020})\BibitemShut {NoStop}%
\bibitem [{\citenamefont {Chmiela}\ \emph {et~al.}(2019)\citenamefont
  {Chmiela}, \citenamefont {Sauceda}, \citenamefont {Poltavsky}, \citenamefont
  {M{\"u}ller},\ and\ \citenamefont {Tkatchenko}}]{sGDMLsoftware2019}%
  \BibitemOpen
  \bibfield  {author} {\bibinfo {author} {\bibfnamefont {S.}~\bibnamefont
  {Chmiela}}, \bibinfo {author} {\bibfnamefont {H.~E.}\ \bibnamefont
  {Sauceda}}, \bibinfo {author} {\bibfnamefont {I.}~\bibnamefont {Poltavsky}},
  \bibinfo {author} {\bibfnamefont {K.-R.}\ \bibnamefont {M{\"u}ller}}, \ and\
  \bibinfo {author} {\bibfnamefont {A.}~\bibnamefont {Tkatchenko}},\ }\href
  {\doibase https://doi.org/10.1016/j.cpc.2019.02.007} {\bibfield  {journal}
  {\bibinfo  {journal} {Comput. Phys. Commun.}\ }\textbf {\bibinfo {volume}
  {240}},\ \bibinfo {pages} {38 } (\bibinfo {year} {2019})}\BibitemShut
  {NoStop}%
\bibitem [{\citenamefont {Doerr}\ \emph {et~al.}(2021)\citenamefont {Doerr},
  \citenamefont {Majewski}, \citenamefont {P\'erez}, \citenamefont {Kr\"amer},
  \citenamefont {Clementi}, \citenamefont {No\'e}, \citenamefont {Giorgino},\
  and\ \citenamefont {De~Fabritiis}}]{Doerr_TorchMD_JCTC2021}%
  \BibitemOpen
  \bibfield  {author} {\bibinfo {author} {\bibfnamefont {S.}~\bibnamefont
  {Doerr}}, \bibinfo {author} {\bibfnamefont {M.}~\bibnamefont {Majewski}},
  \bibinfo {author} {\bibfnamefont {A.}~\bibnamefont {P\'erez}}, \bibinfo
  {author} {\bibfnamefont {A.}~\bibnamefont {Kr\"amer}}, \bibinfo {author}
  {\bibfnamefont {C.}~\bibnamefont {Clementi}}, \bibinfo {author}
  {\bibfnamefont {F.}~\bibnamefont {No\'e}}, \bibinfo {author} {\bibfnamefont
  {T.}~\bibnamefont {Giorgino}}, \ and\ \bibinfo {author} {\bibfnamefont
  {G.}~\bibnamefont {De~Fabritiis}},\ }\href {\doibase
  10.1021/acs.jctc.0c01343} {\bibfield  {journal} {\bibinfo  {journal} {J.
  Chem. Theory Comput.}\ }\textbf {\bibinfo {volume} {17}},\ \bibinfo {pages}
  {2355–2363} (\bibinfo {year} {2021})}\BibitemShut {NoStop}%
\bibitem [{\citenamefont {Sch{\"u}tt}\ \emph {et~al.}(2019)\citenamefont
  {Sch{\"u}tt}, \citenamefont {Kessel}, \citenamefont {Gastegger},
  \citenamefont {Nicoli}, \citenamefont {Tkatchenko},\ and\ \citenamefont
  {M{\"u}ller}}]{SchNetPack}%
  \BibitemOpen
  \bibfield  {author} {\bibinfo {author} {\bibfnamefont {K.~T.}\ \bibnamefont
  {Sch{\"u}tt}}, \bibinfo {author} {\bibfnamefont {P.}~\bibnamefont {Kessel}},
  \bibinfo {author} {\bibfnamefont {M.}~\bibnamefont {Gastegger}}, \bibinfo
  {author} {\bibfnamefont {K.~A.}\ \bibnamefont {Nicoli}}, \bibinfo {author}
  {\bibfnamefont {A.}~\bibnamefont {Tkatchenko}}, \ and\ \bibinfo {author}
  {\bibfnamefont {K.-R.}\ \bibnamefont {M{\"u}ller}},\ }\href {\doibase
  10.1021/acs.jctc.8b00908} {\bibfield  {journal} {\bibinfo  {journal} {J.
  Chem. Theo. Comp.}\ }\textbf {\bibinfo {volume} {15}},\ \bibinfo {pages}
  {448} (\bibinfo {year} {2019})}\BibitemShut {NoStop}%
\bibitem [{\citenamefont {Sauceda}\ \emph {et~al.}(2019)\citenamefont
  {Sauceda}, \citenamefont {Chmiela}, \citenamefont {Poltavsky}, \citenamefont
  {M{\"u}ller},\ and\ \citenamefont {Tkatchenko}}]{sGDML_Appl_jcp}%
  \BibitemOpen
  \bibfield  {author} {\bibinfo {author} {\bibfnamefont {H.~E.}\ \bibnamefont
  {Sauceda}}, \bibinfo {author} {\bibfnamefont {S.}~\bibnamefont {Chmiela}},
  \bibinfo {author} {\bibfnamefont {I.}~\bibnamefont {Poltavsky}}, \bibinfo
  {author} {\bibfnamefont {K.-R.}\ \bibnamefont {M{\"u}ller}}, \ and\ \bibinfo
  {author} {\bibfnamefont {A.}~\bibnamefont {Tkatchenko}},\ }\href {\doibase
  10.1063/1.5078687} {\bibfield  {journal} {\bibinfo  {journal} {J. Chem.
  Phys.}\ }\textbf {\bibinfo {volume} {150}},\ \bibinfo {pages} {114102}
  (\bibinfo {year} {2019})}\BibitemShut {NoStop}%
\bibitem [{\citenamefont {Dormand}\ and\ \citenamefont
  {Prince}(1980)}]{dormand1980family}%
  \BibitemOpen
  \bibfield  {author} {\bibinfo {author} {\bibfnamefont {J.~R.}\ \bibnamefont
  {Dormand}}\ and\ \bibinfo {author} {\bibfnamefont {P.~J.}\ \bibnamefont
  {Prince}},\ }\href {\doibase 10.1016/0771-050X(80)90013-3} {\bibfield
  {journal} {\bibinfo  {journal} {J. of Comp. and Appl. Math.}\ }\textbf
  {\bibinfo {volume} {6}},\ \bibinfo {pages} {19} (\bibinfo {year}
  {1980})}\BibitemShut {NoStop}%
\bibitem [{\citenamefont {Pascanu}\ \emph {et~al.}(2013)\citenamefont
  {Pascanu}, \citenamefont {Mikolov},\ and\ \citenamefont
  {Bengio}}]{pascanu2013difficulty}%
  \BibitemOpen
  \bibfield  {author} {\bibinfo {author} {\bibfnamefont {R.}~\bibnamefont
  {Pascanu}}, \bibinfo {author} {\bibfnamefont {T.}~\bibnamefont {Mikolov}}, \
  and\ \bibinfo {author} {\bibfnamefont {Y.}~\bibnamefont {Bengio}},\ }in\
  \href {\doibase arXiv:1211.5063} {\emph {\bibinfo {booktitle} {Int. Conf.
  Mach. Learn.}}}\ (\bibinfo {organization} {PMLR},\ \bibinfo {year} {2013})\
  pp.\ \bibinfo {pages} {1310--1318}\BibitemShut {NoStop}%
\bibitem [{\citenamefont {Vaswani}\ \emph {et~al.}(2017)\citenamefont
  {Vaswani}, \citenamefont {Shazeer}, \citenamefont {Parmar}, \citenamefont
  {Uszkoreit}, \citenamefont {Jones}, \citenamefont {Gomez}, \citenamefont
  {Kaiser},\ and\ \citenamefont {Polosukhin}}]{vaswani2017attention}%
  \BibitemOpen
  \bibfield  {author} {\bibinfo {author} {\bibfnamefont {A.}~\bibnamefont
  {Vaswani}}, \bibinfo {author} {\bibfnamefont {N.}~\bibnamefont {Shazeer}},
  \bibinfo {author} {\bibfnamefont {N.}~\bibnamefont {Parmar}}, \bibinfo
  {author} {\bibfnamefont {J.}~\bibnamefont {Uszkoreit}}, \bibinfo {author}
  {\bibfnamefont {L.}~\bibnamefont {Jones}}, \bibinfo {author} {\bibfnamefont
  {A.~N.}\ \bibnamefont {Gomez}}, \bibinfo {author} {\bibfnamefont
  {{\L}.}~\bibnamefont {Kaiser}}, \ and\ \bibinfo {author} {\bibfnamefont
  {I.}~\bibnamefont {Polosukhin}},\ }in\ \href
  {https://proceedings.neurips.cc/paper/2017/file/3f5ee243547dee91fbd053c1c4a845aa-Paper.pdf}
  {\emph {\bibinfo {booktitle} {Advances in Neural Information Processing
  Systems}}}\ (\bibinfo {year} {2017})\ pp.\ \bibinfo {pages}
  {5998--6008}\BibitemShut {NoStop}%
\bibitem [{\citenamefont {Bahdanau}\ \emph {et~al.}(2014)\citenamefont
  {Bahdanau}, \citenamefont {Cho},\ and\ \citenamefont
  {Bengio}}]{bahdanau2014neural}%
  \BibitemOpen
  \bibfield  {author} {\bibinfo {author} {\bibfnamefont {D.}~\bibnamefont
  {Bahdanau}}, \bibinfo {author} {\bibfnamefont {K.}~\bibnamefont {Cho}}, \
  and\ \bibinfo {author} {\bibfnamefont {Y.}~\bibnamefont {Bengio}},\
  }\href@noop {} {\  (\bibinfo {year} {2014})},\ \Eprint
  {http://arxiv.org/abs/1409.0473} {arXiv:1409.0473} \BibitemShut {NoStop}%
\bibitem [{\citenamefont {Wu}\ \emph {et~al.}(2016)\citenamefont {Wu},
  \citenamefont {Schuster}, \citenamefont {Chen}, \citenamefont {Le},
  \citenamefont {Norouzi}, \citenamefont {Macherey}, \citenamefont {Krikun},
  \citenamefont {Cao}, \citenamefont {Gao}, \citenamefont {Macherey},
  \citenamefont {Klingner}, \citenamefont {Shah}, \citenamefont {Johnson},
  \citenamefont {Liu}, \citenamefont {Łukasz Kaiser}, \citenamefont {Gouws},
  \citenamefont {Kato}, \citenamefont {Kudo}, \citenamefont {Kazawa},
  \citenamefont {Stevens}, \citenamefont {Kurian}, \citenamefont {Patil},
  \citenamefont {Wang}, \citenamefont {Young}, \citenamefont {Smith},
  \citenamefont {Riesa}, \citenamefont {Rudnick}, \citenamefont {Vinyals},
  \citenamefont {Corrado}, \citenamefont {Hughes},\ and\ \citenamefont
  {Dean}}]{GoogleNMT}%
  \BibitemOpen
  \bibfield  {author} {\bibinfo {author} {\bibfnamefont {Y.}~\bibnamefont
  {Wu}}, \bibinfo {author} {\bibfnamefont {M.}~\bibnamefont {Schuster}},
  \bibinfo {author} {\bibfnamefont {Z.}~\bibnamefont {Chen}}, \bibinfo {author}
  {\bibfnamefont {Q.~V.}\ \bibnamefont {Le}}, \bibinfo {author} {\bibfnamefont
  {M.}~\bibnamefont {Norouzi}}, \bibinfo {author} {\bibfnamefont
  {W.}~\bibnamefont {Macherey}}, \bibinfo {author} {\bibfnamefont
  {M.}~\bibnamefont {Krikun}}, \bibinfo {author} {\bibfnamefont
  {Y.}~\bibnamefont {Cao}}, \bibinfo {author} {\bibfnamefont {Q.}~\bibnamefont
  {Gao}}, \bibinfo {author} {\bibfnamefont {K.}~\bibnamefont {Macherey}},
  \bibinfo {author} {\bibfnamefont {J.}~\bibnamefont {Klingner}}, \bibinfo
  {author} {\bibfnamefont {A.}~\bibnamefont {Shah}}, \bibinfo {author}
  {\bibfnamefont {M.}~\bibnamefont {Johnson}}, \bibinfo {author} {\bibfnamefont
  {X.}~\bibnamefont {Liu}}, \bibinfo {author} {\bibnamefont {Łukasz Kaiser}},
  \bibinfo {author} {\bibfnamefont {S.}~\bibnamefont {Gouws}}, \bibinfo
  {author} {\bibfnamefont {Y.}~\bibnamefont {Kato}}, \bibinfo {author}
  {\bibfnamefont {T.}~\bibnamefont {Kudo}}, \bibinfo {author} {\bibfnamefont
  {H.}~\bibnamefont {Kazawa}}, \bibinfo {author} {\bibfnamefont
  {K.}~\bibnamefont {Stevens}}, \bibinfo {author} {\bibfnamefont
  {G.}~\bibnamefont {Kurian}}, \bibinfo {author} {\bibfnamefont
  {N.}~\bibnamefont {Patil}}, \bibinfo {author} {\bibfnamefont
  {W.}~\bibnamefont {Wang}}, \bibinfo {author} {\bibfnamefont {C.}~\bibnamefont
  {Young}}, \bibinfo {author} {\bibfnamefont {J.}~\bibnamefont {Smith}},
  \bibinfo {author} {\bibfnamefont {J.}~\bibnamefont {Riesa}}, \bibinfo
  {author} {\bibfnamefont {A.}~\bibnamefont {Rudnick}}, \bibinfo {author}
  {\bibfnamefont {O.}~\bibnamefont {Vinyals}}, \bibinfo {author} {\bibfnamefont
  {G.}~\bibnamefont {Corrado}}, \bibinfo {author} {\bibfnamefont
  {M.}~\bibnamefont {Hughes}}, \ and\ \bibinfo {author} {\bibfnamefont
  {J.}~\bibnamefont {Dean}},\ }\href@noop {} {\  (\bibinfo {year} {2016})},\
  \Eprint {http://arxiv.org/abs/arXiv:1609.08144} {arXiv:1609.08144}
  \BibitemShut {NoStop}%
\bibitem [{\citenamefont {Sauceda}\ \emph
  {et~al.}(2020{\natexlab{a}})\citenamefont {Sauceda}, \citenamefont {Chmiela},
  \citenamefont {Poltavsky}, \citenamefont {M{\"u}ller},\ and\ \citenamefont
  {Tkatchenko}}]{sgdml_bookAppl}%
  \BibitemOpen
  \bibfield  {author} {\bibinfo {author} {\bibfnamefont {H.~E.}\ \bibnamefont
  {Sauceda}}, \bibinfo {author} {\bibfnamefont {S.}~\bibnamefont {Chmiela}},
  \bibinfo {author} {\bibfnamefont {I.}~\bibnamefont {Poltavsky}}, \bibinfo
  {author} {\bibfnamefont {K.-R.}\ \bibnamefont {M{\"u}ller}}, \ and\ \bibinfo
  {author} {\bibfnamefont {A.}~\bibnamefont {Tkatchenko}},\ }in\ \href
  {https://koreauniv.pure.elsevier.com/en/publications/construction-of-machine-learned-force-fields-with-quantum-chemica}
  {\emph {\bibinfo {booktitle} {Construction of Machine Learned Force Fields
  with Quantum Chemical Accuracy: Applications and Chemical Insights}}}\
  (\bibinfo {year} {2020})\ pp.\ \bibinfo {pages} {277--307}\BibitemShut
  {NoStop}%
\bibitem [{\citenamefont {Sauceda}\ \emph
  {et~al.}(2020{\natexlab{b}})\citenamefont {Sauceda}, \citenamefont
  {Gastegger}, \citenamefont {Chmiela}, \citenamefont {M{\"u}ller},\ and\
  \citenamefont {Tkatchenko}}]{sauceda2020JCP}%
  \BibitemOpen
  \bibfield  {author} {\bibinfo {author} {\bibfnamefont {H.~E.}\ \bibnamefont
  {Sauceda}}, \bibinfo {author} {\bibfnamefont {M.}~\bibnamefont {Gastegger}},
  \bibinfo {author} {\bibfnamefont {S.}~\bibnamefont {Chmiela}}, \bibinfo
  {author} {\bibfnamefont {K.-R.}\ \bibnamefont {M{\"u}ller}}, \ and\ \bibinfo
  {author} {\bibfnamefont {A.}~\bibnamefont {Tkatchenko}},\ }\href {\doibase
  10.1063/5.0023005} {\bibfield  {journal} {\bibinfo  {journal} {J. Chem.
  Phys.}\ }\textbf {\bibinfo {volume} {153}},\ \bibinfo {pages} {124109}
  (\bibinfo {year} {2020}{\natexlab{b}})}\BibitemShut {NoStop}%
\bibitem [{\citenamefont {Sch{\"u}tt}\ \emph {et~al.}(2021)\citenamefont
  {Sch{\"u}tt}, \citenamefont {Unke},\ and\ \citenamefont {Gastegger}}]{painn}%
  \BibitemOpen
  \bibfield  {author} {\bibinfo {author} {\bibfnamefont {K.}~\bibnamefont
  {Sch{\"u}tt}}, \bibinfo {author} {\bibfnamefont {O.}~\bibnamefont {Unke}}, \
  and\ \bibinfo {author} {\bibfnamefont {M.}~\bibnamefont {Gastegger}},\ }in\
  \href {https://proceedings.mlr.press/v139/schutt21a.html} {\emph {\bibinfo
  {booktitle} {Int. Conf. Mach. Learn.}}}\ (\bibinfo  {publisher} {PMLR},\
  \bibinfo {year} {2021})\ pp.\ \bibinfo {pages} {9377--9388}\BibitemShut
  {NoStop}%
\end{thebibliography}%

\begin{acknowledgments}
KRM was supported in part by the Institute of Information \& Communications Technology Planning \& Evaluation (IITP) grant funded by the Korea Government (No. 2019-0-00079,  Artificial Intelligence Graduate School Program, Korea University), and was partly supported by the German Ministry for Education and Research (BMBF) under Grants 01IS14013A-E, 01GQ1115, 01GQ0850, 01IS18025A and 01IS18037A; the German Research Foundation (DFG) under Grant Math+, EXC 2046/1, Project ID 390685689.  
HES works at the BASLEARN - TU Berlin/BASF Joint Lab for Machine Learning, co-financed by TU Berlin and BASF SE.
Correspondence should be addressed to HES and KRM.
\end{acknowledgments}

\end{document}